\documentclass[sn-mathphys,Numbered]{sn-jnl}

\usepackage{graphicx}%
\usepackage{multirow}%
\usepackage{amsmath,amssymb,amsfonts}%
\usepackage{amsthm}%
\usepackage{mathrsfs}%
\usepackage[title]{appendix}%
\usepackage{xcolor}%
\usepackage{textcomp}%
\usepackage{manyfoot}%
\usepackage{booktabs}%
\usepackage{algorithm}%
\usepackage{algorithmicx}%
\usepackage{algpseudocode}%
\usepackage{listings}%

\usepackage{lineno}
\usepackage{upgreek}

\theoremstyle{thmstyleone}%
\theoremstyle{thmstyletwo}%

\theoremstyle{thmstylethree}%

\begin{document}

\title{Breakdown of Conventional Winding Number Calculation in One-Dimensional Lattices with Interactions Beyond Nearest Neighbors}
\author[1]{\fnm{Amir} \sur{Rajabpoor Alisepahi}}

\author[2]{\fnm{Siddhartha} \sur{Sarkar}}

\author[2]{\fnm{Kai} \sur{Sun}}

\author*[1,3,4]{\fnm{Jihong} \sur{Ma}}\email{Jihong.Ma@uvm.edu}

\affil*[1]{\orgdiv{Department of Mechanical Engineering}, \orgname{University of Vermont}, \orgaddress{\street{33 Colchester Ave}, \city{Burlington}, \postcode{05405}, \state{VT}, \country{USA}}}

\affil[2]{\orgdiv{Department of Physics}, \orgname{University of Michigan}, \orgaddress{\street{450 Church St.}, \city{Ann Arbor}, \postcode{48109}, \state{MI}, \country{USA}}}

\affil[3]{\orgdiv{Department of Physics}, \orgname{University of Vermont}, \orgaddress{\street{33 Colchester Ave}, \city{Burlington}, \postcode{05405}, \state{VT}, \country{USA}}}

\affil[4]{\orgdiv{Materials Science Program}, \orgname{University of Vermont}, \orgaddress{\street{33 Colchester Ave}, \city{Burlington}, \postcode{05405}, \state{VT}, \country{USA}}}

\abstract{Topological insulators hold promises to realize exotic quantum phenomena in electronic, photonic, and phononic systems. Conventionally, topological indices, such as winding numbers, have been used to predict the number of topologically protected domain-wall states (TPDWSs) in topological insulators, a signature of the topological phenomenon called bulk-edge correspondence. Here, we demonstrate theoretically and experimentally that the number of TPDWSs in a mechanical Su-Schrieffer-Heeger (SSH) model can be higher than the winding number depending on the strengths of beyond-nearest-neighbor interactions, revealing the breakdown of the winding number prediction. Alternatively, we resort to the Berry connection to accurately characterize the number and spatial features of TPDWSs in SSH systems, further confirmed by the Jackiw-Rebbi theory proving that the multiple TPDWSs correspond to the bulk Dirac cones. Our findings deepen the understanding of complex network dynamics and offer a generalized paradigm for precise TPDWS prediction in potential applications involving localized vibrations, such as drug delivery and quantum computing.}

\maketitle

\section{Introduction}\label{sec1}

As a special class of mechanical metamaterials and phononic crystals, topological mechanical metamaterials and phononic crystals endowed with anomalous wave manipulation capabilities have attracted significant attention over the past decade. Analogous to topological insulators in quantum physics~\cite{haldane1988model, kane2005quantum, hasan2010colloquium,qi2011topological}, where a topological invariant is introduced to classify different quantum states of matter, in mechanical systems, such a topological invariant can also be derived from the spectral evolution of eigenvectors, or mode shapes, from a unit cell analysis to determine the number and types of topologically protected surface/edge/corner states confining phonon modes both statically~\cite{kane2014topological, paulose2015topological, rocklin2017transformable, rocklin2016mechanical, stenull2016topological, bilal2017intrinsically} and dynamically~\cite{ma2018edge,ma2019valley,susstrunk2015observation, nash2015topological, wang2015topological, mousavi2015topologically, kariyado2015manipulation, pal2016helical, brendel2017pseudomagnetic, chaunsali2018subwavelength, prodan2017dynamical,luo2021observation,wang2021elastic,ni2019observation,qi2020acoustic,zhang2019second,chen2021corner}, usually referred to as the bulk-edge correspondence.

One illustrative example of employing a topological invariant to determine the nontrivial topologically protected domain-wall states (TPDWSs) can be seen in the one-dimensional (1D) Su-Schrieffer-Heeger (SSH) model~\cite{su1979solitons,su1980soliton} as shown in Fig. S1a, b in Supplementary Note 1. Initially introduced to study solitons in polyacetylene, the SSH model was later adapted in mechanical systems to identify TPDWSs via a winding number calculation~\cite{lubensky2015phonons,esmann2018topological,pal2017edge}. As discussed in the references above and the ``Analysis of One-Dimensional Su-Schrieffer-Heeger Model" subsection in Methods, the two arrangements of isomers with different spring stiffness $c_1$ and $c_2$ in Fig. S1b represent two topologically distinct phases. When $c_1>c_2$, the origin is excluded in the contour plot in the complex plane of the off-diagonal term in the stiffness matrix, $\mathbf{C}(k)$, where $k$ is the wave number in the reciprocal space, and thus, the winding number $n=0$, signifying a trivial intra-cell-hopping phase. In contrast, when $c_1<c_2$, the contour winds about the origin once, $i.e.$, $n=1$, indicating a topologically nontrivial inter-cell-hopping phase, see Fig. S1c in Supplementary Note 1. These gauge-dependent winding numbers can also be evaluated via the Zak phase~\cite{zak1989berry} measuring the rotation of eigenvectors in the unit cell, see Fig. S2 in Supplementary Note 1. Alternatively, the winding number can be directly calculated from $\mathbf{C}(k)$: 
\begin{eqnarray}
	n=\int_{-\pi/a}^{\pi/a}\frac{1}{4\pi i}\mathrm{tr}[\boldsymbol{\upsigma}_3\mathbf{C}^{\prime-1}\partial_k\mathbf{C}^{\prime}]dk,
	\label{WN}
\end{eqnarray}
where $\boldsymbol{\upsigma}_3$ is the third Pauli matrix, and $\mathbf{C}'$ is similar to an effective Hamiltonian, which is a chiral matrix obtained from:
\begin{eqnarray}
	\mathbf{C}'(k)=\mathbf{C}(k)-(c_1+c_2)\boldsymbol{\upsigma}_0,
\end{eqnarray}
where $\boldsymbol{\upsigma}_0$ is the identity matrix. Seaming two phases with different $n$s creates a domain wall where a localized mode emerges. Such a domain-wall state is topologically protected due to the intrinsic topological phase difference between the two domains, $i.e.$, the bulk-edge correspondence.

The above discussion has been well understood and applied to systems of higher dimensions, such as the 2D quantum valley Hall effect in phononic crystals~\cite{pal2017edge, ma2019valley,lu2016valley,lu2017observation,liu2018tunable,liu2019experimental,pal2016helical}. Most of these studies can be simplified using mass-spring systems considering only the nearest neighbor (NN) interactions. Recently, arising attention has been devoted to mechanical metamaterials with lattice interactions beyond nearest neighbors (BNNs), achieving roton-like acoustic dispersion relations under ambient conditions similar to those observed in correlated quantum systems at low temperatures~\cite{chen2021roton,iglesias2021experimental,iorio2022roton,cui2022tunable,wang2022nonlocal,zhu2022observation}. In addition, intriguing topological states also arise due to such BNN coupling, including the increased winding number corresponding to a higher number of edge states due to larger BNN differences, as reported in previous studies~\cite{grundmann2020topological,chen2018study,liu2023acoustic}. 

In this study, we report the impact of BNN couplings on bulk-edge correspondence in addition to the increased winding number. Despite being commonly believed that the number of TPDWSs is governed by the difference in winding numbers between the two domains, we prove that the number of TPDWSs in the SSH model is not dictated by the winding number, but by the Jackiw-Rebbi (JR) indices associated with the JR zero modes~\cite{jackiw1976solitons}. In previously studied SSH models, the winding numbers and JR indices happened to predict the same number of TPDWSs. However, our investigation reveals that such a coincidence is not generic, $i.e.$, in the presence of BNN couplings, these two indices can significantly deviate from each other. In such a generic setup, we prove analytically and verify numerically and experimentally, that the JR indices always correctly predict the number of TPDWSs, while the relationship between the winding number and TPDWSs generally fails. We note that the discrepancy between the two is not a special scenario in an SSH model but a rather generic phenomenon across all topological Maxwell lattices and chiral matters \cite{guzman2022geometry}.
We, thus, propose to use the Berry connection with distinguishable local winding numbers as an alternative topological index to identify TPDWs, which also applies to a broader range of lattices with BNN interactions beyond those presented in the main text. 

\section{Results and Discussion}\label{sec2}

\subsection{Mass-Spring Model Analysis}\label{subsec2}

\begin{figure} [h!]
		\centering
		\textbf{Fig. 1 Unit-cell analysis of the Su-Schrieffer-Heeger model with identical third-nearest neighbors.}
	\includegraphics[scale=0.48]{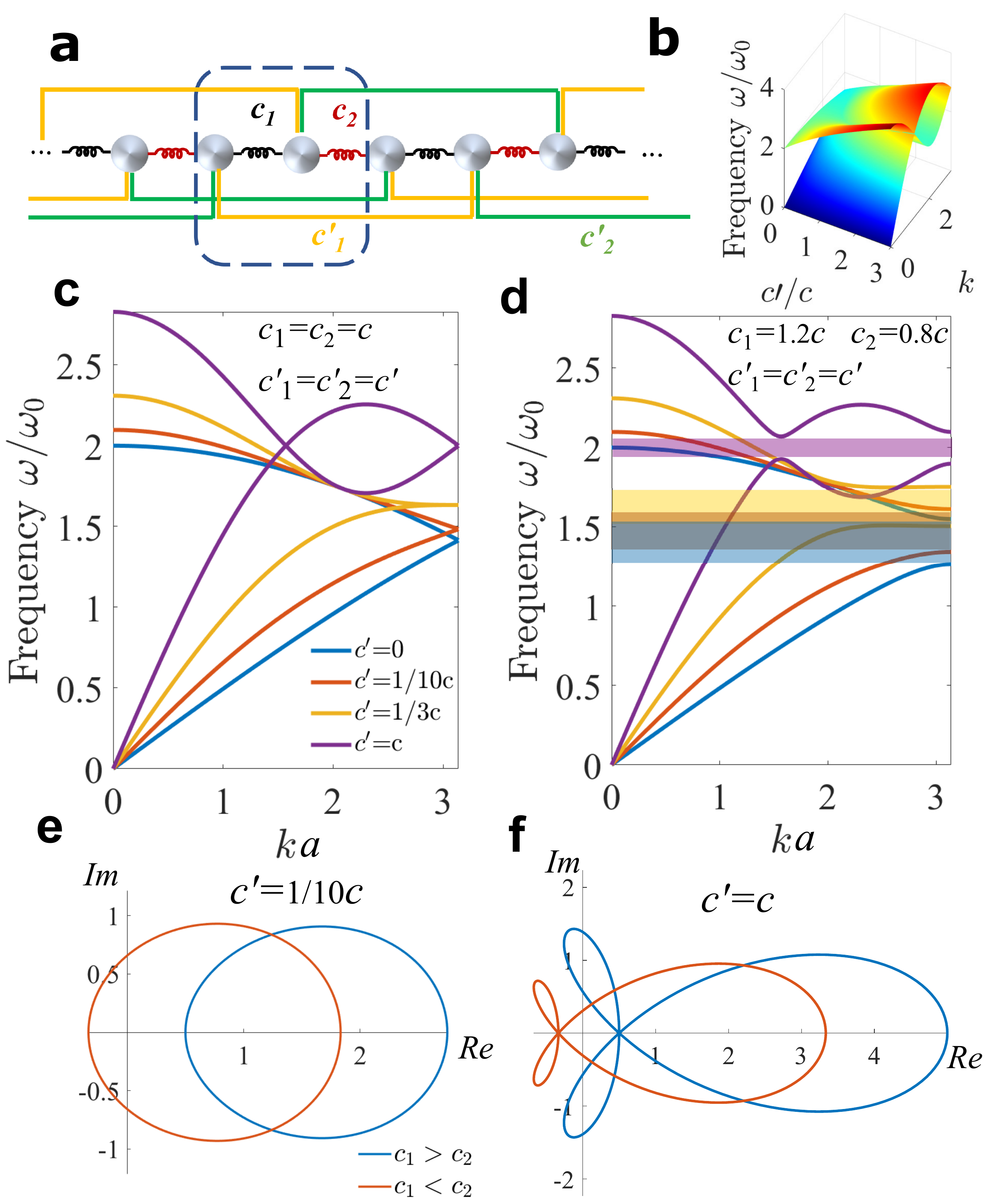} 
	\caption{\textbf{a} Unit cell (circled in a dashed line) of a chain of identical masses, $m$, with nearest neighbors (NNs) with spring constants $c_1$ (black springs) and $c_2$ (red springs), and third-nearest neighbors (TNNs) with spring constants $c^\prime_1$ (gold lines) and $c^\prime_2$ (green lines), respectively. \textbf{b} 3D representation of phonon band variation with identical NN stiffness $c$ and TNN strength $c'$ (with lattice spacing $a=1$). $\omega_0$=$\sqrt{c/m}$. \textbf{c} and \textbf{d} Unit cell band structures with various $c^\prime$ and \textbf{c} identical and \textbf{d} non-identical $c$. Shaded areas in \textbf{d} denote the bandgaps between acoustic and optical phonon branches with matching colors. \textbf{e} and \textbf{f} Contour plots of the off-diagonal element of the chiral matrix, $\mathbf{C}'(k)$, in the complex plane for a complete circuit of $k$ from $k=0$ to $2\pi$ for unit cells with \textbf{e} $c^\prime=1/10c$ and \textbf{f} $c^\prime=c$.}
	\label{unitcell_toy}
\end{figure}

To start with, we add third-nearest neighbors (TNNs) with spring stiffness $c'$ to a 1D mass-spring chain of lattice spacing $a$ with a NN spring stiffness $c$ and restrict the motion of identical masses to the horizontal direction, as presented in Fig.~\ref{unitcell_toy}a. The phonon dispersion of a pair of masses reveals that when $c'<1/3c$, the acoustic and optical phonon bands cross at $k=\pi/a$, protected by the space inversion symmetry (SIS). When $c'>1/3c$, two additional band crossings emerge in the irreducible Brillouin zone (IBZ), as presented in Fig.~\ref{unitcell_toy}b, c. Derivation of the exact locations of the Dirac points due to the existence of the TNN is presented in Eqns.~\ref{spring_mass_3NN_eqn1}-\ref{Dirac_points} in the ``Analysis of One-Dimensional Su-Schrieffer-Heeger Model with Third-Nearest Neighbors" subsection in Methods. The additional band folding due to strong $c'$s results in negative group velocities in the acoustic phonon branch, corresponding to the backward wave observed in the previous study~\cite{chen2021roton}. We then break the SIS by applying a small perturbation to the NN spring stiffness $c$, $i.e.$, making $c_1=0.8c$ and $c_2=1.2c$, while maintaining all TNNs identical, which opens a band gap between acoustic and optical bands, as shown in Fig.~\ref{unitcell_toy}d. The winding number calculation from $\mathbf{C}'(k)$ in Eqn.~\ref{chiral_TNN} for the two isomers of such a system suggests that, regardless of the strength of $c'$, the difference between two phases is always one, indicating one TPDWS at the domain boundary of the two phases. Note that, with the existence of $c'$, the contour plots in the complex plane are no longer circular as those shown in Fig. S1c. With weak $c'$ (for example, $c'=1/10c$), they present oval shapes, as shown in Fig.~\ref{unitcell_toy}e, while strong $c'$ (such as $c'=c$) creates two additional loops along the path, Fig.~\ref{unitcell_toy}f. In either case, the circuit winds around the origin exactly once when $c_1<c_2$, indicating the topological charge being $n=1$, while excluding it when $c_1>c_2$, thus $n=0$, yielding a consistent $n$ difference. 

\begin{figure} [h!]
	\centering
	\textbf{Fig. 2 Supercell analysis of the Su-Schrieffer-Heeger model with a domain wall with weak and strong identical third-nearest neighbors.}
	\includegraphics[scale=0.1]{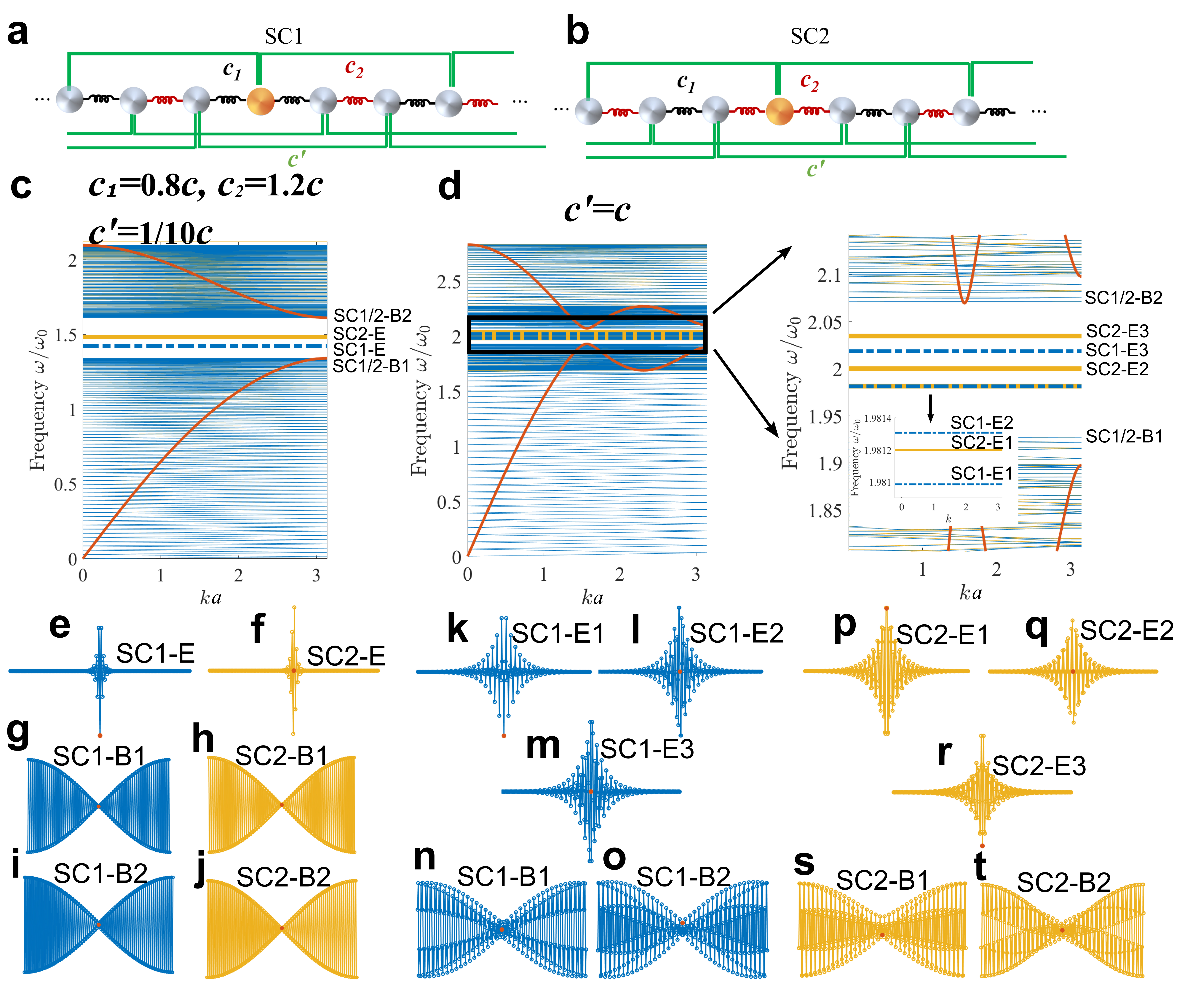} 
	\caption{\textbf{a} and \textbf{b} Supercells featuring the two arrangements of nearest neighbors (NNs) with \textbf{a} soft-SC1 and \textbf{b} stiff-SC2 springs with spring constants $c_1=0.8c$ and $c_2=1.2c$, respectively, where $c$ is an arbitrary spring constant, connected to the domain-wall mass (highlighted in orange). \textbf{c} and \textbf{d} Band diagrams of the supercells with \textbf{c} weak (with a spring constant $c'=1/10c$) and \textbf{d} strong ($c'=c$) third NNs (TNNs), respectively. Blue and yellow bands correspond to supercell \textbf{a} and \textbf{b}, respectively. Red curves are bulk bands acquired from the unit cell analysis in Fig.~\ref{unitcell_toy}. Phonon bands within the bulk bandgap for the case of $c'=c$ in \textbf{d} are zoomed in next to the complete dispersion. Dashed blue and bold solid yellow bands in \textbf{c} and \textbf{d} are domain-wall modes, denoted as SC1-E(1-3) and SC2-E(1-3), respectively. Bands below and above the bulk bandgap are marked as SC1/2-B1/2. The domain-wall and bulk mode shapes with $c'=1/10c$ ($c'=c$)  are presented in \textbf{e}-\textbf{j} (\textbf{k}-\textbf{t}) with blue and yellow colors matching SC1 and SC2, respectively. Red solid circles in \textbf{e}-\textbf{t} denote the displacements of domain-wall masses. Although visualized in the vertical directions, all mass displacements are $de$ $facto$ in the horizontal direction.}
	\label{supercell_toy}
\end{figure}

To verify the number of TPDWSs predicted with winding numbers, we consider a $supercell$ containing 301 masses with a domain-wall mass at the center connected to soft or stiff springs on both sides, as shown in  Fig.~\ref{supercell_toy}a, b, about which are the symmetrically arranged two phases with different $n$s. Bloch conditions are applied at the two ends of the chain to mimic an infinitely long chain for phonon dispersion calculation. Details of the Bloch conditions for the supercell are presented in the ``Supercell Analysis of the Su-Schrieffer-Heeger Model" subsection in Methods. As predicted, when connected by weak TNNs, $i.e.$, $c'=1/10c$, only one symmetric (asymmetric) TPDWS exists within the bulk bandgap when the domain-wall mass is connected by soft (stiff) NN springs, as shown in the supercell band structures in Fig.~\ref{supercell_toy}c with corresponding edge and bulk mode shapes presented in Fig.~\ref{supercell_toy}e-j. However, when $c'=c$, we identify two additional edge modes in the bulk bandgap, Fig.~\ref{supercell_toy}d, violating the aforementioned winding number prediction. Mode shapes of these emerging bands in the bulk bandgap shown in Fig.~\ref{supercell_toy}k-m, p-r, confirm the localization of displacements at the domain wall, distinguishable from the bulk modes, as shown in Fig.~\ref{supercell_toy}n,o,s,t. When the domain-wall mass is connected by two soft (stiff) springs $c_1=0.8c$ ($c_2=1.2c$), we obtain one (two) symmetric and two (one) asymmetric displacement fields about the domain-wall mass, as can be seen in Fig.~\ref{supercell_toy}k-m (p-r).

Spatial Fourier transform (SFT) of the mode shapes presented in Fig.~\ref{JRmodes} show a significantly widened peak width in all these edge modes compared to their bulk counterparts, suggesting a faster spatial decay of the vibration from the domain wall, evident of an edge mode. It is worth noting that, due to strong TNN interactions, the additional band crossing at $k=\pi/2$ as shown in Fig.~\ref{unitcell_toy}c when $c_1=c_2$ (the location of which is expressed in Eqn.~\ref{Dirac_points} in the ``Analysis of One-Dimensional Su-Schrieffer-Heeger Model with Third-Nearest Neighbors" subsection in Methods) results in the global peak (valley) of the bulk acoustic (optical) mode occurring at $k=\pi/2$, instead of $k=\pi$ where a local peak (valley) appears, as in Fig.~\ref{supercell_toy}d. Hence, sharper SFT peaks appear at $k=\pi/2$ in the bulk modes closest to the bandgap ($i.e.$, SC1/2-B1/2), as shown in Fig.~\ref{JRmodes}b. In-gap TPDWSs are presented as widened peaks located at $k=\pi/2$, $\pi$, and $3\pi/2$, as shown in Figs. ~\ref{JRmodes}b and S3 in Supplementary Note 2, indicating their rapid spatial decay away from the domain wall with hybridized wavelengths.

\subsection{Jackiw Rebbi Zero Modes}\label{subsec2}
\begin{figure} [h!]
	\centering
	\textbf{Fig. 3 Spatial Fourier transform of supercell mode shapes in lattices with weak and strong identical third-nearest neighbors.}
	\includegraphics[scale=0.1]{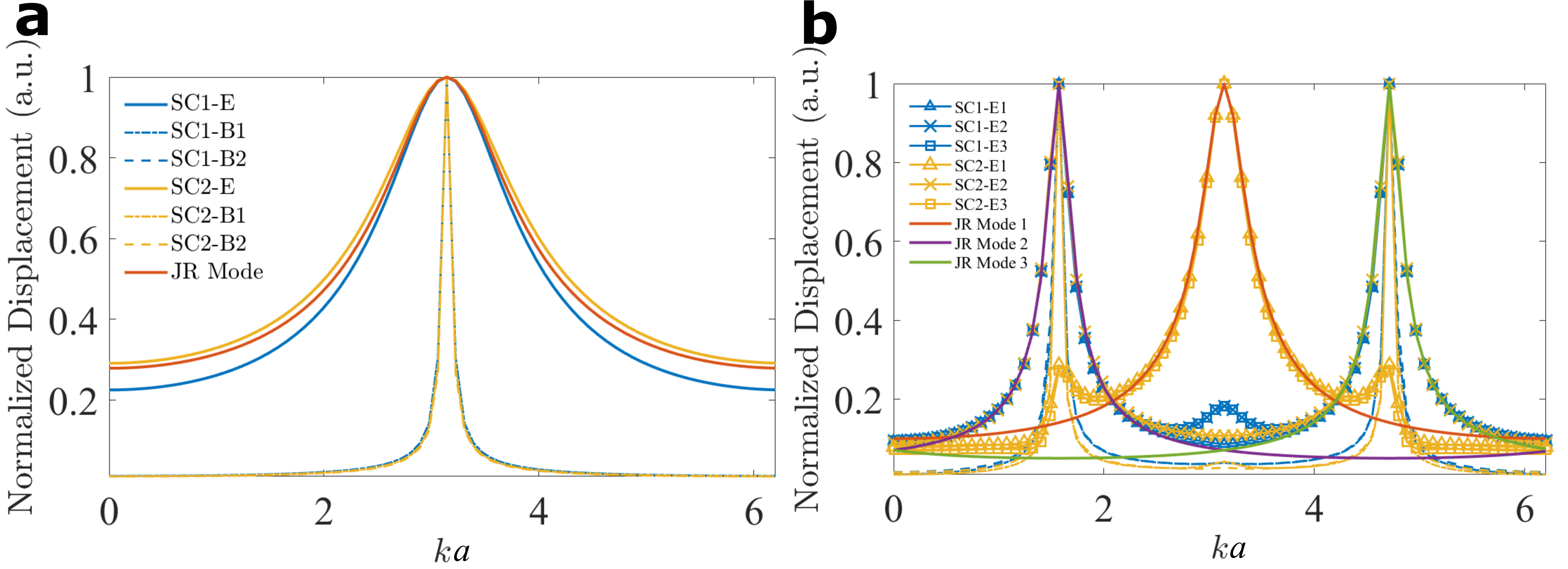} 
	\caption{\textbf{a} Spatial Fourier transform (SFT) of mode shapes of the lattice with weak third-nearest neighbors (TNNs) presented in Fig.~\ref{supercell_toy}e (blue solid), f (yellow solid), g (blue dot-dash), h (yellow dot-dash), i (blue dash), and j (yellow dash).  The SFT of Jackiw Rebbi (JR) zero mode expressed in Eqn.~\ref{JR_pi_smalleta} in Methods is plotted as the red solid curve. \textbf{b} SFT of mode shapes of the lattice with strong TNNs presented in Fig. ~\ref{supercell_toy}k (blue triangle), l (blue x), m (blue square), n (blue dot-dash), o (blue dash), p (yellow triangle), q (yellow x), r (yellow square), s (yellow dot-dash), and t (yellow dash). The SFT of the three JR modes corresponding to strong TNNs expressed in Eqns.~\ref{JR_pi_largeeta}  and ~\ref{JR_other_sol} in Methods are expressed as red (Eqn.~\ref{JR_pi_largeeta}), purple (Eqn.~\ref{JR_other_sol} with the``$+$'' sign in the solution) and green (Eqn.~\ref{JR_other_sol} with the ``$-$'' sign in the solution) solid curves. Note that the SFT plots in \textbf{b} indicate the spatial decay starting from the domain-wall mass. Plots starting from the mass right next to the domain wall are presented in Fig. S3 in Supplementary Note 2.}
	\label{JRmodes}
\end{figure}

  These domain-wall states can also be characterized by a massless Dirac theory. For $c_1>c_2$ ($c_1<c_2$), the breaking of the SIS introduces a positive (negative) mass to each Dirac point. Due to the mass sign flipping at the domain wall, one TPDWS in the bandgap, known as the \textit{JR zero mode}, is expected to arise at the domain wall for each Dirac cone~\cite{jackiw1976solitons}. Thus, the number of crossings presented in Fig. ~\ref{unitcell_toy}c within the IBZ equals that of TPDWSs when the bandgap is open. Such an agreement also strongly resembles those in the quantum valley Hall effect in 2D, where the number of in-gap TPDWSs also matches that of bulk Dirac cones \cite{ma2019valley}. A comprehensive demonstration of the existence of TPDWSs due to the hybridization of JR modes corresponding to the three Dirac points within the IBZ and their analytical solutions characterizing the spatial decay are presented in the ``Derivation of the Jackiw Rebbi Zero Modes" subsection in Methods. As presented in Figs.~\ref{JRmodes}, and S3, SFT of the analytical JR zero modes match well with those obtained from the supercell analysis, indicating successful prediction of TPDW properties using the JR theory, as well as the breakdown of the winding number prediction. 

In principle, regardless of the strengths of $c'$, the analytical solutions of JR modes in the SSH model should all stay at $\omega/\omega_0=\sqrt{(c_1+c_2+2c')/m}$ (or $\omega^2/\omega_0^2=0$ if plotting the eigenvalues of $\mathbf{C}'(k)$ in Eqn.~\ref{chiral_TNN}). However, this is not the case if we simply flip the arrangement of springs about the domain-wall mass as presented in Fig.~\ref{supercell_toy}a, b. One reason is that the domain wall setup breaks the chiral symmetry of the stiffness matrix $\mathbf{C}(k)$ since the total spring constants about the domain-wall mass is $2c_1+2c'$ ($2c_2+2c'$) for SC1 (SC2), different from $c_1+c_2+2c'$ around other masses in the supercell. Such a perturbation can be readily fixed by pinning the interface to the ground using an additional spring with a constant of $c_2-c_1$ ($c_1-c_2$) for SC1 (SC2), as shown in Fig.S4a (b) in Supplementary Note 2. The two TPDWSs due to different domain walls for weak TNNs (for example, $c'=1/10c$) then become degenerate at $\omega/\omega_0=\sqrt{(c_1+c_2+2c')/m}$, or $\omega^2/\omega_0^2=0$ if removing the diagonal elements of $\mathbf{C}(k)$, as shown in Fig.S4c. However, for the ones with strong TNNs (for example, $c'=c$) where multiple TPDWSs exist at one domain wall, such a fix can only bring one of the TPDWSs to $\omega/\omega_0=\sqrt{(c_1+c_2+2c')/m}$ (or $\omega^2/\omega_0^2=0$ if removing the diagonal elements of $\mathbf{C}(k)$). The other two TPDWSs are located symmetrically above and below it, as shown in Fig. S4d. The remaining shift in frequency is due to the hybridization of two JR modes with the same parity. Details about TPDWS symmetry and parity, as well as their hybridization conditions causing the frequency shift, are all discussed in ``Jackiw Rebbi Mode Parity and Hybridization" subsection in Methods. The shifting of the zero-frequency(energy) domain-wall modes (after pinning the domain-wall mass) to finite frequencies is not unique to the SSH model, instead, it is a generic feature existing in most known 0D topological modes in various types of topological insulators (such as corner modes in 2D higher-order topological insulators), whose energies are also sensitive to local perturbations near the localized modes\cite{van2020topological,proctor2020robustness}.
\subsection{Berry Connection}\label{subsec2}
\begin{figure} [h!]
	\centering
	\textbf{Fig. 4 Berry connection variation with third-nearest neighbor strength. }
	\includegraphics[scale=0.27]{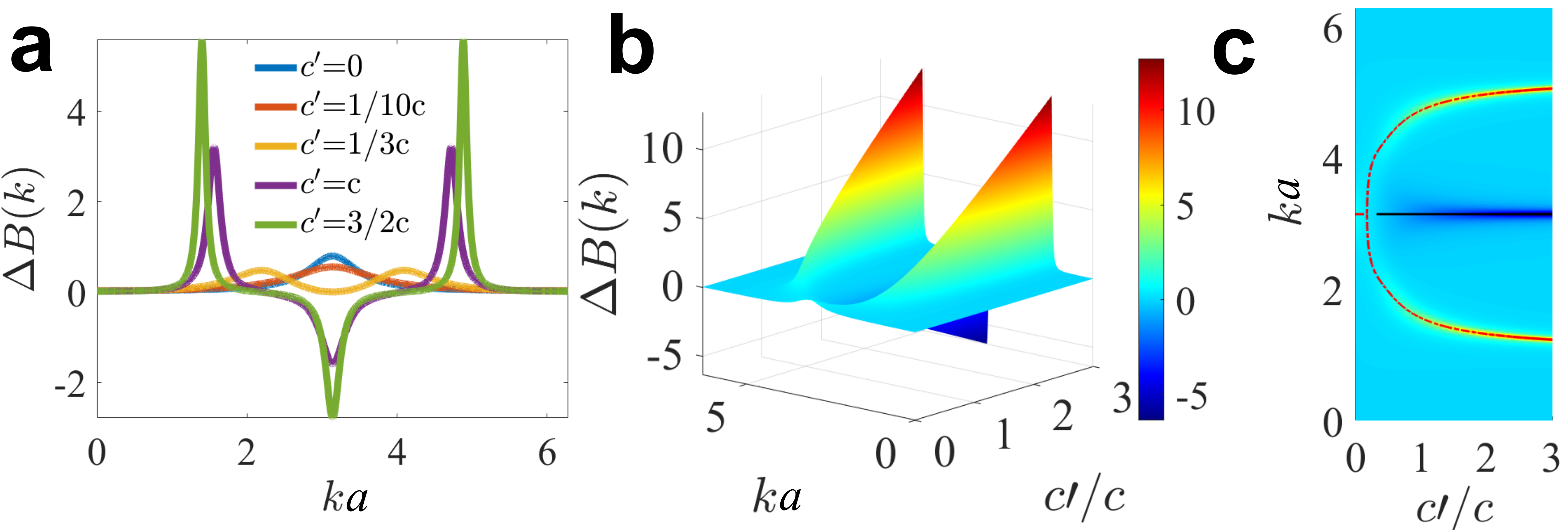} 
	\caption{\textbf{a} Winding number difference, $\Delta B(k)$, from $ka=0$ to $2\pi$ (where $k$ is the wave number and $a$ is the lattice spacing) in the irreducible Brillouin zone with different third-nearest neighbor strengths, $c'$. \textbf{b} and \textbf{c} 3D visualization of the evolution of $\Delta B(k)$ with $c'$ ($c$ is the strength of nearest neighbors). The red and black curves in the top view shown in \textbf{c} indicate peaks and valleys of the Berry connections, respectively.}
	\label{WN_diff}
\end{figure}

The question then arises as to how to determine TPDWSs using a topological descriptor associating the spectral evolution of the eigenvector with these states. A closer examination of the contour plots in Fig.~\ref{unitcell_toy}e, f and the winding number calculation in Eqn.~\ref{WN} suggest that, although the difference in $n$ is one regardless of the TNN strength, trajectories of the contour plots, or the integrand of Eqn.~\ref{WN}, $i.e.$, the Berry connection,
\begin{eqnarray}
	B(k)=\frac{1}{4\pi i}\mathrm{tr}[\boldsymbol{\upsigma}_3\mathbf{C}^{\prime-1}\partial_k\mathbf{C}^{\prime}],
	\label{BC}
\end{eqnarray} 
varies with $c'$, where $\mathbf{C}'$ is expanded to Eqn.~\ref{chiral_TNN} in the ``Analysis of One-Dimensional Su-Schrieffer-Heeger Model with Third-Nearest Neighbors" in Methods to include the TNNs. Since the number of TPDWSs depends on the topological invariant difference due to different gauges, we plot $\Delta B(k)=B_1(k)-B_2(k)$, where $B_1(k)$ [$B_2(k)$] refers to the case when $c_1<c_2$ ($c_1>c_2$), for unit cells with different $c'$ in Fig.~\ref{WN_diff} to describe its topology. When $c'=0$, only one peak exists at $k=\pi/a$ in $\Delta B$, corresponding to the Dirac point at $k=\pi/a$ in the band structure in Fig.~\ref{unitcell_toy}c. As $c'$ increases while $c'<1/3c$, this peak at $k=\pi/a$ decreases and widens until it splits into two smaller peaks (such as when $c'=1/3c$). As $c'$ continues to increase, the valley at $k=\pi/a$ dips below $\Delta B(k)=0$ while the two positive peaks drift apart with locations matching Dirac points as expressed in Eqn.~\ref{Dirac_points}, until $k=\pi/3a$ and $5\pi/3a$, as discussed in the ``Analysis of One-Dimensional Su-Schrieffer-Heeger Model with Third Nearest Neighbors" in Methods. Meanwhile, the two peaks and one valley are further sharpened as $c'$ increases. The integral around each peak (valley) is $\pm$1, $i.e.$, yielding a local winding number. It is worth noting that the total integral over the IBZ does not change as $c'$ varies, yielding a consistent winding number of $n=1$. The transition from one peak in $\Delta B$ into two peaks and one valley agrees with the change of TPDWS counts with corresponding $c'$. Moreover, the locations of the peaks/valleys informing of the TPDWS wavelengths also agree with those calculated from JR zero modes demonstrated in ``Derivation of the Jackiw Rebbio Zero Modes" and the supercell calculation presented in ``Supercell Analysis of the Su-Schrieffer-Heeger Model" in Methods, the results of which are plotted in Figs.~\ref{JRmodes}, S3, and S4u, v in Supplementary Note 2 if fixing the domain wall to make $\mathbf{C}(k)$ chiral. 

To understand such a transition from one peak to two peaks and one valley with increased $c'$, one can draw an analogy between the evolution of $\Delta B$ as $c'$ decreases and the inter-valley mixing of the Berry curvature in our previously studied valley Hall effect \cite{ma2019valley}. In the current SSH model with TNNs, when $c'\gg c$, the perturbation induced by the NNs, $c$, is relatively small, resulting in minimal inter-valley mixing between the two peaks and one valley in $\Delta B$, which distinctively exist in the IBZ, matching the three TPDWSs within the bulk bandgap. As $c'$ weakens, the difference in $c$ becomes more prominent, introducing stronger SIS perturbation, and thus an enhanced peak-valley mixing closer to the valley at $\pi/a$, and eventually merging all into one single peak at $k=\pi/a$, leaving only one TPDWS in the bandgap. 

It is worth noting that, compared to conventional winding number calculations, the Berry connection prediction alluded to is not limited to making correct TPDWS predictions in lattices with identical TNNs. As discussed in subsection ``Topologically Protected Domain-Wall States beyond Equal Third-Nearest Neighbors" in Methods, one can also predict the number and the wavelengths of TPDWS when TNNs are nonidentical, $i.e.$, $c_1'\neq c_2'$, as well as for systems with interactions beyond TNNs, whose results are shown in Figs. S5 and S6, respectively, in Supplementary Note 3. The additional TPDWs due to BNNs always arise in pairs with local integrals of Berry connection around the peaks and valleys being $\pm 1$, respectively. Information about wavelengths acquired from such a Berry connection analysis is also unattainable using conventional winding number calculations. Hence, the Berry connection provides a generalized methodology supplying enriched information about TPDWSs in lattices with complex networks. 

\subsection{Laser-Assisted Experimental Characterization}\label{subsec2}
\begin{figure} [h!]
\centering
\textbf{Fig. 5 Experimental characterization of topological domain-wall states in lattices with strong and weak third-nearest neighbors. }
	\includegraphics[scale=0.45]{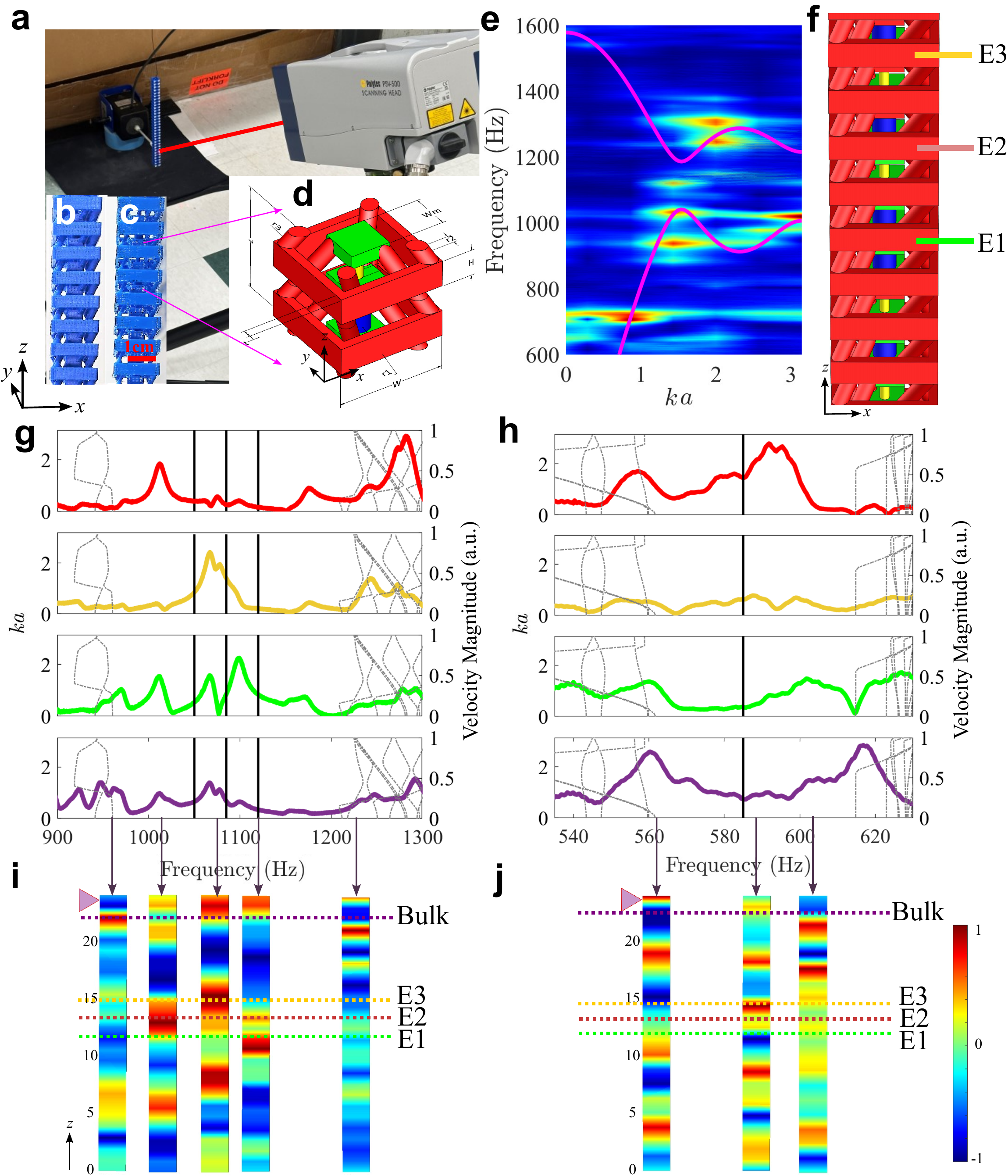} 
	\caption{\textbf{a} Experimental setup showing the scanning laser Doppler vibrometer and the shaker attached to the specimen bottom. \textbf{b} and \textbf{c} Zoomed-in views of the specimens with \textbf{b} strong and \textbf{c} weak third-nearest neighbors (TNNs), with a unit cell in \textbf{d}, where masses (green blocks) connected by alternating stiff (thick blue bars) and soft (thin yellow bars) springs are also linked by red frames and bars as TNNs, whose strength is adjustable by tuning the red bar diameter. All unit cell dimensions are in the ``Experimental Fabrication and Characterization" subsection in Methods. \textbf{e} 2D map of the discrete Fourier transform of experimental data matching the torsional phonon bands from the unit cell analysis (magenta-solid). \textbf{f} Zoomed-in view around the domain-wall mass connected by two blue bars. E1, E2, and E3 indicate the frames near the domain wall where torsional velocities are measured and presented as green, red, and yellow curves (with the velocity magnitude axis on the right) in \textbf{g} (strong TNNs) and \textbf{h} (weak TNNs), where bulk and domain-wall modes from the supercell analysis are also plotted as dotted and solid curves (with the vertical axis showing the wavenumber, $ka$, on the left), respectively. \textbf{i} and \textbf{j} Torsional velocity fields excited (triangle near the top) at frequencies denoted by arrows above for lattices with \textbf{i} strong and \textbf{j} weak TNNs. Dashed lines with matching bar colors in \textbf{f} denote the locations of three TNN frames. The purple dashed lines near the excitation points are bulk regions, whose frequency responses are shown as purple curves in \textbf{g} and \textbf{h}. Torsional velocity profiles in \textbf{e}, \textbf{i}, and \textbf{j} are all normalized by the highest magnitude with a color bar shown in \textbf{j}. }
	\label{experiment}
\end{figure}

We proceed now to conduct experiments on 1D specimens adapted from an existing TNN model~\cite{chen2021roton,iglesias2021experimental}, as shown in Fig. \ref{experiment}a-d. Information regarding the experimental specimens is listed in subsection ``Experimental Fabrication and Characterization" in Methods. As presented in Fig.~\ref{experiment}d, each unit cell contains a pair of masses connected by alternating stiff and soft NN bars and identical TNN frames. The domain-wall mass behind the frame labeled as E1 in Fig.~\ref{experiment} f is connected by two stiff struts (blue bars), about which are placed with 8 unit cells with opposite stiff and soft NN arrangements. The lattice specimens are hung by a string from the top and are excited in the $y$-direction using an electrodynamic shaker (PCB 2007E01, powered by a Krohn-Hite 7500 amplifier) placed off-centered near the bottom left end for torsional excitation. Velocities of the left and right ends of each frame in Fig.~\ref{experiment}b, c in the $y-$direction are measured by a scanning laser Doppler vibrometer (SLDV, Polytec PSV-500), and their differences are recorded as the torsional velocities about the $z$-axis. Note that shear deformations in the $y$-direction will also be recorded simultaneously. However, these shear modes do not exist in our frequency range of interest. For comparison, unit cell and supercell analyses containing a domain wall with the same dimensions and material parameters are also performed using the finite element method with COMSOL Multiphysics with results presented in Fig. S7 in Supplementary Note 4.   \\ 

We then prescribe a chirp excitation sweeping from 800 to 1400 Hz to the lattice with strong TNNs, Fig. \ref{experiment}b (500 to 700 Hz to the one with weak TNNs, Fig.~\ref{experiment}c) and measure torsional velocities. For the specimen with strong TNNs and without a domain wall, we achieve an excellent agreement between the spatiotemporal spectral response obtained from a discrete Fourier transform of the torsional velocity sampled along the axial direction of the specimen ($i.e.$, the $z$-axis in Fig.~\ref{experiment}) and the acoustic and optical torsional phonon branches predicted from the unit cell analysis, as shown in Fig.~\ref{experiment}e, presenting a roton-like dispersion relation. Frequency responses of the experimentally measured torsional velocities of the frames near the domain wall and in the bulk of the specimen with strong (weak) TNNs reveal three (one) distinct peaks within the bulk bandgap with amplified torsional velocities in proximity to the domain wall, as evident in Fig.~\ref{experiment}g (h). These in-gap peaks agree with the TPDWSs predicted in supercell band structures containing matching lattice configurations with Bloch boundary conditions applied at two ends, presented as dark solid lines in Fig.~\ref{experiment}g, h. Snapshots of the torsional velocity fields experimentally measured at bulk and TPDWS frequencies are shown in Fig.~\ref{experiment}i and j, corresponding to strong and weak TNNs, respectively, agreeing with mode shapes calculated using the finite element method plotted in Fig. S7 in Supplementary Note 4. Symmetries of these measured and calculated TPDWSs concur with those in the toy model presented in Fig.~\ref{supercell_toy}p-r and f, \textit{i.e.}, the three torsional TPDWSs in the lattice with strong TNNs investigated the experimentally measured Fig.~\ref{experiment}i, simulated Fig. S7a, and the mass-spring model in Fig.~\ref{supercell_toy}p-r, are asymmetric, symmetric, and symmetric about the domain wall (corresponding to the red, yellow, and green peaks in Fig.~\ref{experiment}g), and the one TPDWS in the lattice with weak TNNs shown in Figs.~\ref{experiment}j, S7b, and \ref{supercell_toy}f are all asymmetric (corresponding to the red peak in Fig.~\ref{experiment}h), confirming that these domain-wall modes are indeed TPDWSs predicted in theory. 

\section{Conclusions}\label{sec3}
We theoretically and experimentally reveal the breakdown of the conventional winding number prediction of TPDWSs in SSH lattices with BNN interactions. We, instead, propose to count the local winding numbers by calculating the Berry connection characterizing the evolution of eigenvectors in the reciprocal space to obtain the correct number of TPDWSs. Moreover, Berry connection offers more insights into the TPDWSs, including their wavelengths and spatial decay rates. Further, we demonstrate that these TPDWSs are the phonon realization of JR zero modes, analytically validating the Berry connection prediction. Note that the discordance between the total winding number in IBZ and the counting of topological modes isn't an exception; rather, it is a common characteristic universally seen across topological states governed by winding numbers. For example, a similar discrepancy is also evident in topological Maxwell lattices and chiral matters\cite{guzman2022geometry}, where the accurate enumeration of topological modes requires the winding number to be adjusted by the addition of integer numbers. Such amendments encapsulate the intricate aspects of the system's physics, such as lattice structures, gauge choice, and local counting. Furthermore, our study provides a more generalized paradigm in accurate topological state predictions in lattices beyond 1D with TNNs, and is applicable to a broader range of complex systems with multi-nodal interactions \cite{sun2012surface,prodan2017dynamical}, especially at the nano-\cite{prodan2009topological} and microscales \cite{ma2023phonon}, where BNN interactions commonly exist. Successfully identifying and achieving mechanical/vibration topological states in complex systems can also inspire solutions in other realms of science where predictable and precise vibration modes are critical, such as facilitating drug delivery \cite{vanden2019raman,yeo2010ultrasonic} and advancing quantum information processing using phonons \cite{bienfait2019phonon,chen2020entanglement,wang2019hexagonal} in quantum technologies, such as quantum computing.

\section{Methods}\label{sec4}
\subsection{Analysis of One-Dimensional Su-Schrieffer-Heeger Model}
The governing equations of a 1D SSH lattice unit cell shown in Fig. S1 in Supplementary Note 1 can be expressed as:
\begin{eqnarray}
	m\ddot{u}_1^n=c_1(u_2^n-u_1^n)-c_2(u_1^n-u_2^{n-1}),\label{spring_mass_eqn1}\\
	m\ddot{u}_2^n=c_2(u_1^{n+1}-u_2^n)-c_1(u_2^n-u_1^n),
	\label{spring_mass_eqn2}
\end{eqnarray}
where displacements of the two masses in the $n$-th cell are denoted as $u_1^n$ and $u_2^n$, respectively, and can be expressed using a plane-wave solution in combination with Bloch-Floquet periodic boundary conditions:
\begin{eqnarray}
	{\mathbf u}^n(t)={\tilde{\mathbf{u}}}(k)e^{i(nka-\omega t)},
	\label{planewave}
\end{eqnarray}
where $\omega$ is the vibration frequency, ${\mathbf u}^n$ are the displacements of the $n$-th cell with $\mathbf{u}^n=[u_1^n,u_2^n]$, $k$ is the wave number, which is inversely proportional to the wavelength $\lambda$, \textit{i.e.}, $k=2\pi/\lambda$, $a$ denotes the lattice constant, $\tilde{\mathbf{u}}(k)$ are displacements within the unit cell. Substituting this expression in Eqns.~\ref{spring_mass_eqn1} and~\ref{spring_mass_eqn2} gives:
\begin{eqnarray}
	[\mathbf{C}(k)-\omega^2m]{\tilde{\mathbf{u}}}(k)=0,
	\label{goveqn}
\end{eqnarray}
where $\mathbf{C}(k)$ is the stiffness matrix of the periodic system:
\begin{equation}
	\mathbf{C}(k)=
	\begin{bmatrix}
		c_1+c_2 & -c_1-c_2e^{-ika}\\
		-c_1-c_2e^{ika} & c_1+c_2 \\
	\end{bmatrix},
	\label{stiffness}
\end{equation}
Assume $a=1$ and divide the stiffness matrix $\mathbf{C}(k)$ by $c_2$, we can then plot the off-diagonal element of $\mathbf{C}(k)$, $i.e$., $\rho(k)=c_1/c_2+e^{ika}$, and project $\rho(k)$ to a complex plane, as shown in Fig. S1c in Supplementary Note 1. 

The winding number difference between the two gauges can also be characterized by the Zak phase:
\begin{eqnarray}
	Z=\frac{i}{\pi}\int_{-\pi/a}^{\pi/a}{\tilde{\mathbf u}}_-^*(k)\partial_k{\tilde{\mathbf u}}_-(k)dk,
	\label{Zak}
\end{eqnarray}
where ${\tilde{\mathbf u}}_-(k) = [\rho^*(k)/|\rho(k)|,1]/\sqrt{2}$ is the eigenvector corresponding to the smaller eigenvalue of the matrix $\mathbf{C}(k)$. Writing $\rho(k) = |\rho(k)| e^{i\phi(k)}$, we find $Z=\frac{i}{2\pi}\int_{-\pi/a}^{\pi/a} dk\, \partial_k (-i \phi(k))=\frac{1}{2\pi}\int_{-\pi/a}^{\pi/a} dk\, \partial_k \phi(k)$. This implies that the Zak phase measures the change in the phase of the first component (in this particular choice of gauge) of the eigenvector as wavenumber $k$ changes from $-\pi/a$ to $\pi/a$.  When $c_1>c_2$, $Z=0$, suggesting no changes in the phase difference of the eigenvectors across IBZ, see the blue curve in Figs. S1c and S2a. When $c_1<c_2$, $Z=1$, indicating a phase change of $2\pi$ within the IBZ, see the red curve in Figs. S1c and S2b in Supplementary Note 1. 

\subsection{Analysis of One-Dimensional Su-Schrieffer-Heeger Model with Third-Nearest Neighbors}
Adding TNNs, as in Fig.~\ref{unitcell_toy}a, modifies the governing equations to:
\begin{eqnarray}
	m\ddot{u}_1^n=c_1(u_2^n-u_1^n)-c_2(u_1^n-u_2^{n-1})+c_1'(u_2^{n+1}-u_1^n)-c_2'(u_1^n-u_2^{n-2}),\label{spring_mass_3NN_eqn1}\\
	m\ddot{u}_2^n=c_2(u_1^{n+1}-u_2^n)-c_1(u_2^n-u_1^n)+c_2'(u_1^{n+2}-u_2^{n})-c_1'(u_2^{n}-u_1^{n-1}).
	\label{spring_mass_3NN_eqn2}
\end{eqnarray}
Plugging the same wave ansatz in Eqn.~\eqref{planewave} in the above set of equations, we obtain an equation of the same form as Eqn.~\eqref{goveqn} with the following stiffness matrix:
\begin{equation}
	\mathbf{C}(k)=
	\begin{bmatrix}
		c_1+c_2 +c_1'+c_2' & -c_1-c_2e^{-ika}-c_1'e^{ika}-c_2' e^{-2ika}\\
		-c_1-c_2e^{ika}-c_1'e^{-ika}-c_2' e^{2ika}& c_1+c_2 +c_1'+c_2'  \\
	\end{bmatrix}.
	\label{stiffness_3NN}
\end{equation}
The chiral matrix then becomes:
\begin{eqnarray}
	\mathbf{C}'(k)=\mathbf{C}(k)-(c_1+c_2+c_1'+c_2')\boldsymbol{\upsigma}_0,
	\label{chiral_TNN}
\end{eqnarray}
The contour plots of the off-diagonal element of this matrix $\rho(k) = c_1+c_2e^{ika}+c_1'e^{-ika}+c_2' e^{2ika}$ for a complete circuit of $k$ from $k=0$ and $k=2\pi/a$ for different values of TNN spring stiffness, $c_1'=c_2'=c'$, are plotted in Fig.~\ref{unitcell_toy}e, f. We see that for $c_1<c_2$ ($c_1>c_2$) the winding number of $\rho(k)$ around the origin is one (zero).

It is also instructive to find the values of $k$ at which the Dirac points appear when $c_1=c_2=c$ and $c_1'=c_2' = \eta c$. To get this, we note that the band gap closes when the off-diagonal term in the matrix $\mathbf{C}(k)$ in Eqn.~\ref{stiffness_3NN} is zero:
\begin{equation}
	c+\frac{c}{z}+\eta c z +\frac{\eta c}{z^2} = 0 \Rightarrow z = -1, \frac{-1+\eta \pm \sqrt{(1+\eta)(1-3\eta)}}{2\eta},
	\label{Dirac_eq}
\end{equation}
where $z = e^{ika}$. Clearly, there are three Dirac points when $\left|\frac{-1+\eta \pm \sqrt{(1+\eta)(1-3\eta)}}{2\eta}\right|=1$ since $z = e^{ika}$. Now, there are two cases: (i) $\frac{-1+\eta \pm \sqrt{(1+\eta)(1-3\eta)}}{2\eta}$ is real, (ii) $\frac{-1+\eta \pm \sqrt{(1+\eta)(1-3\eta)}}{2\eta}$ is complex. In the first case, we have:
\begin{equation}
	\begin{split}
		&\left(-1+\eta \pm \sqrt{(1+\eta)(1-3\eta)}\right)^2 = 4\eta^2 \\
		\Rightarrow & \eta = -1 \text{ or }\eta = 1/3 \text{ or } \eta = 0,\\
		\label{Dirac_sol1}
	\end{split}
\end{equation}
The solution of $z$ in Eqn.~\ref{Dirac_eq} prevents $\eta$ from being zero since it can blow up $z$. A negative $\eta$ is also not possible because there is no negative spring constant. Therefore, the only valid solution is $\eta = 1/3$, resulting in all three Dirac points appearing at $z=-1$, $i.e.$, $k =\pi/a$. 

On the other hand, if $\frac{-1+\eta \pm \sqrt{(1+\eta)(1-3\eta)}}{2\eta}$ is complex, the requirement of the existence of three Dirac points is:
\begin{equation}
	\begin{split}
		&\left(-1+\eta + \sqrt{(1+\eta)(1-3\eta)}\right)\left(-1+\eta - \sqrt{(1+\eta)(1-3\eta)}\right) = 4\eta^2 \\
		\Rightarrow & 1-2\eta +\eta^2 - 1+2\eta+3\eta^2 = 4\eta^2,
		\label{Dirac_sol2}
	\end{split}
\end{equation}
which is always true. Thus, if $\frac{-1+\eta \pm \sqrt{(1+\eta)(1-3\eta)}}{2\eta}$ is complex, its absolute value is always 1, regardless of the value of $\eta$. Note that this condition is only valid when $\sqrt{(1+\eta)(1-3\eta)}$ is complex, meaning $\eta>1/3$ (since $\eta\geq0$). Hence, combining the complex and real solutions of $z$, we can identify the three Dirac points for $\eta\geq 1/3$. The values of $k$ at which these three Dirac points appear are:
\begin{equation}
	k = \frac{\pi}{a}, \pm \frac{1}{a}\arctan\left(\frac{\sqrt{(1+\eta)(3\eta-1)}}{-1+\eta}\right).
	\label{Dirac_points}
\end{equation}
Note that in the limit of $c' \gg c$ or $\eta\rightarrow \infty$, the three Dirac points appear at $k = \pi/a$ and $\pm\pi/3a$ (the latter of which are equivalent to $\pi/3a$, and $5\pi/3a$ from 0 to $2\pi/a$), corresponding to the band crossing locations presented in Fig.~\ref{unitcell_toy}c.
\subsection{Supercell Analysis of the Su-Schrieffer-Heeger Model}
Let's consider supercell SC1 in Fig.~\ref{supercell_toy}a. To obtain the phonon dispersion and mode shapes of the supercell with a domain wall, Bloch boundary conditions are applied at the two ends of the supercell containing 301 masses to mimic an infinitely large lattice with periodic domain walls 300 masses apart. 
	The governing equation for each mass in the supercell is as follows:
	\begin{eqnarray}
		m\ddot{u}_p^n=c_1(u_{p+1}^n-u_p^n )+c_2 (u_{p-1}^n-u_p^n )+c'(u_{p+3}^n-u_p^n )+c'(u_{p-3}^n-u_p^n ),\label{supercell_eqn}\\
		m\ddot{u}_p^n=c_1(u_{p+2}^n-u_{p+1}^n )+c_2 (u_{p}^n-u_{p+1}^n )+c'(u_{p+4}^n-u_{p+1}^n )+c'(u_{p-2}^n-u_{p+1}^n ),\label{supercell_eqn2}
	\end{eqnarray}
	where the subscript denotes the $pth$ mass is the supercell and the superscript denotes the $nth$ cell. Below are the governing equations of displacements of the three beginning masses at the left end, $u_1$, $u_2$, and $u_3$, and the three ending masses at the right end, $u_{P}$, $u_{P-1}$, and $u_{P-2}$, where $P$ is the total number of masses, which equals 301 in our model, within the current supercell, $n$:\\ 
	\begin{eqnarray}
		m\ddot{u}_1^n=c_1(u_{2}^n-u_1^n )+c_2 (u_{P}^{n-1}-u_1^n )+c'(u_{4}^n-u_1^n )+c'(u_{P-2}^{n-1}-u_1^n ),\label{supercell_eqn3}\\
		m\ddot{u}_2^n=c_2(u_{3}^n-u_2^n )+c_1 (u_{1}^n-u_2^n )+c'(u_{5}^n-u_2^n )+c'(u_{P-1}^{n-1}-u_2^n ),\label{supercell_eqn4}\\
		m\ddot{u}_3^n=c_1(u_{4}^n-u_3^n )+c_2 (u_{2}^n-u_3^n )+c'(u_{6}^n-u_3^n )+c'(u_{P}^{n-1}-u_3^n ),\label{supercell_eqn5}\\
		m\ddot{u}_{P-2}^n=c_2(u_{P-1}^n-u_{P-2}^n )+c_1 (u_{P-3}^n-u_{P-2}^n )+c'(u_{1}^{n+1}-u_{P-2}^n )+c'(u_{P-5}^{n}-u_{P-2}^n ),\label{supercell_eqn6}\\
		m\ddot{u}_{P-1}^n=c_2(u_{P}^n-u_{P-1}^n )+c_1 (u_{P-2}^n-u_{P-1}^n )+c'(u_{2}^{n+1}-u_{P-1}^n )+c'(u_{P-4}^{n}-u_{P-1}^n ),\label{supercell_eqn7}\\
		m\ddot{u}_{P}^n=c_2(u_{1}^{n+1}-u_{P}^n )+c_1 (u_{P-1}^n-u_{P}^n )+c'(u_{3}^{n+1}-u_{P}^n )+c'(u_{P-3}^{n}-u_{P}^n ),\label{supercell_eqn8}
	\end{eqnarray}
 Displacements of supercell $n$ in relation to the $n-1$th and $n+1$th supercells, respectively, are:
	\begin{eqnarray}
		u_1^{n+1}(t)=u_1^{n}(k)e^{i(ka-\omega t)},\label{sc1}\\
		u_2^{n+1}(t)=u_2^{n}(k)e^{-i(ka-\omega t)},\label{sc2}\\
		u_3^{n+1}(t)=u_3^{n}(k)e^{i(ka-\omega t)},\label{sc3}\\
		u_{P}^{n-1}(t)=u_{P}^{n}(k)e^{i(-ka-\omega t) },\label{sc4}\\
		u_{P-1}^{n-1}(t)=u_{P-1}^{n}(k)e^{i(-ka-\omega t)},\label{sc5}\\
		u_{P-2}^{n-1}(t)=u_{P-2}^{n}(k)e^{i(ka-\omega t)},\label{sc6}
	\end{eqnarray}
	Applying these boundary conditions to the displacements of the three masses at two ends of the supercell results in the supercell stiffness matrix $\mathbf{C}(k)$ as:\\
	\begin{equation} 
		\mathbf{C}(k)=\begin{bmatrix}
			c_1+c_2+2c'&-c_1&0 &&&&&-c'e^{-ika}&0&-c_2e^{-ika}\\
			-c_1&c_1+c_2+2c'&-c_2 &&&\hdots&&0&-c'e^{-ika}&0\\
			0&-c_2&c_1+c_2+2c'&&&&&0&0&-c'e^{-ika}\\
			&\vdots&&\ddots&&&&&\vdots&\\        
			&&&-c_2&&2c_2+2c'&-c_2&&&\\
			-c'e^{ika}&0&0&&&&&c_1+c_2+2c'&-c_2&0\\
			0&-c'e^{ika}&0&&&\hdots&&-c_2&c_1+c_2+2c'&-c_1\\
			-c_2e^{ika}&0&-c'e^{ika}&&&&&0&-c_1&c_1+c_2+2c'\\
		\end{bmatrix}
		\label{Supercell_stiffness}
	\end{equation}
	The eigenvalues and eigenvectors of this matrix yield $\omega^2$ and supercell mode shapes, respectively, as shown in Fig. \ref{supercell_toy}.

Since matrix $\mathbf{C}(k)$ in Eqn.~\ref{Supercell_stiffness} is not strictly chiral due to the sum of spring constants at the domain-wall mass is different from $c_1+c_2+2c'$, we add an additional spring with a constant of $c_2-c_1$ ($c_1-c_2$ at the domain-wall mass when it is connected by two soft (stiff) springs, $c_1$ ($c_2$), as shown in Fig. S4a (b) in Supplementary Note 2. We can then plug $\mathbf{C}(k)$ for such a setup into Eqn.~\Ref{chiral_TNN} to obtain $\mathbf{C}'(k)$ with strict chiral symmetry. The eigenvalues ($\omega^2/\omega_0^2$ are henceforth symmetric about $\omega^2/\omega_0^2=0$, as presented in Fig. S4c, d in Supplementary Note 2. Mode shapes of $\mathbf{C}'(k)$ are almost identical to those calculated from $\mathbf{C}(k)$ before adding an additional spring to the domain-wall mass, except that the sequences of the three modes in the case with strong TNNs are reordered, with the two above and below $\omega^2/\omega_0^2=0$ sharing the same parity under inversion symmetry, opposite from the one located at $\omega^2/\omega_0^2=0$, as shown in Fig. S4e-t in Supplementary Note 2.

\subsection{Derivation of the Jackiw Rebbi Zero Modes}
In this section, we demonstrate the existence of multiple TPDWSs as shown in Fig.~\ref{supercell_toy}d in using the JR theory. With an SIS, $i.e.$, $c_1=c_2$, each band crossing point can be characterized by a massless Dirac theory. For $c_1>c_2$ ($c_1<c_2$), the breaking of SIS introduces a positive (negative) mass to each Dirac point. Due to the mass sign flipping at the domain boundary, one TPDWS in the bandgap, known as the JR zero mode, is expected to arise at the domain boundary for each Dirac cone\cite{jackiw1976solitons}, which explains the matching number of TPDWSs with $c_1 \ne c_2$ and band crossing points with $c_1=c_2$. The agreement of the two numbers also strongly resembles those in the quantum valley Hall effect in 2D, where the number of in-gap TPDWSs also matches that of bulk Dirac cones \cite{ma2019valley}. Below, we provide a comprehensive demonstration of the existence of one TPDWS corresponding to each Dirac point and their analytical solutions characterizing the spatial decay observed in both the toy model analysis in Fig.~\ref{supercell_toy} and the experimental observation in Fig.~\ref{experiment}.

\subsubsection{Jackiw Rebbi Mode Corresponding to Dirac Point at $k=\pi/a$} 
Setting $c_1=c+m/2$, $c_2=c-m/2$, and$c_1'=c_2'=\eta c$, and expanding matrix $\mathbf{C}'(k=\pi/a+\delta k)$ for small $\delta k$ and $m$, we get:
\begin{equation}
	\mathbf{C}'(\pi/a+\delta k) \approx c\begin{bmatrix}
		0 & -m - i (1-3\eta) a \delta k\\
		-m + i (1-3\eta) a \delta k & 0
	\end{bmatrix} = -m\boldsymbol{\upsigma}_1 + c(1-3\eta)a \delta k \sigma_2,
\end{equation}
where $\boldsymbol{\upsigma}_i$ are Pauli matrices. In the above equation, $m$ is the mass of the Dirac particle. Now, we create a domain wall at $x=0$ with phase $c_1>c_2$ in the region $x<0$ and $c_1<c_2$ in the region $x>0$. Hence, $m(x)>0$ for $x<0$ and $m(x)<0$ for $x>0$, indicating the location dependence of mass $m$. Since the translation symmetry is broken due to the presence of the domain wall, we can replace $\delta k$ with $-i \partial_x$. We seek a zero frequency domain wall eigenmode $\psi(x)$:
\begin{equation}
	\begin{split}
		&C' \psi(x) = 0\\
		\Rightarrow & (-m(x) \boldsymbol{\upsigma}_1 - c(1-3\eta) a i \boldsymbol{\upsigma}_2 \partial_x) \psi(x) = 0\\
		\Rightarrow & \boldsymbol{\upsigma}_1(-m(x) \boldsymbol{\upsigma}_1 - c(1-3\eta) a i \boldsymbol{\upsigma}_2 \partial_x) \psi(x) = 0\\ 
		\Rightarrow & (-m(x) \mathbf{1} + c(1-3\eta) a \boldsymbol{\upsigma}_3 \partial_x) \psi(x) = 0.
	\end{split}
\end{equation}
Now, there are two cases: (i) $(1-3\eta)>0$, and (ii) $(1-3\eta)<0$. The eigenmodes corresponding to these two scenarios are discussed below.
\begin{enumerate}[(i)]
	\item \underline{$(1-3\eta)>0$:} Plugging the ansatz $\psi(x) = f(x)\begin{pmatrix} 1 \\0 \end{pmatrix}$, where $f(x)$ is a scalar function, we obtain the following differential equation for $f(x)$:
	\begin{equation}
		\partial_x f(x) = \frac{m(x)}{c(1-3\eta)a} \Rightarrow f(x) = c_0 e^{\frac{1}{ca}\int_0^x dx' m(x')/(1-3\eta)},
	\end{equation}
	where $c_0$ is a constant. Note that $f(x)$ decays exponentially away from $x = 0$, since $\frac{m(x>0)}{c(1-3\eta)a}<0$ and $\frac{m(x<0)}{c(1-3\eta)a}>0$. Recall that the zero frequency domain wall mode is at $k=\pi/a$, the full expression of the mode is:
	\begin{equation}
		\psi_{\pi/a}(x) = \psi(x)e^{i \pi x/a} = c_0 e^{\frac{1}{ca}\int_0^x dx' m(x')/(1-3\eta)} e^{i \pi x/a} \begin{pmatrix} 1 \\0 \end{pmatrix}.
		\label{JR_pi_smalleta}
	\end{equation}
	\item \underline{$(1-3\eta)<0$:} Plugging the ansatz $\psi(x) = f(x)\begin{pmatrix} 0 \\1 \end{pmatrix}$, where $f(x)$ is a scalar function, we obtain the following differential equation for $f(x)$:
	\begin{equation}
		\partial_x f(x) = -\frac{m(x)}{c(1-3\eta)a} \Rightarrow f(x) = c_0 e^{-\frac{1}{ca}\int_0^x dx' m(x')/(1-3\eta)}.
	\end{equation}
	Similarly, $f(x)$ decays exponentially away from $x = 0$ since $\frac{m(x>0)}{c(1-3\eta)a}>0$ and $\frac{m(x<0)}{c(1-3\eta)a}<0$. The zero frequency domain wall mode being at $k=\pi/a$ leads to the full expression of the mode being:
	\begin{equation}
		\psi_{\pi/a}(x) = \psi(x)e^{i \pi x/a} = c_0 e^{-\frac{1}{ca}\int_0^x dx' m(x')/(1-3\eta)} e^{i \pi x/a} \begin{pmatrix} 0 \\1 \end{pmatrix}.
		\label{JR_pi_largeeta}
	\end{equation}
\end{enumerate}

\subsubsection{Jackiw Rebbi Mode Corresponding to Dirac Point at $k=\pm \frac{1}{a}\arctan\left(\frac{\sqrt{(1+\eta)(3\eta-1)}}{-1+\eta}\right)$}

For simplicity, we will show here the existence of zero modes for $\eta = 1$ (this is what is considered in Fig.~\ref{supercell_toy}(e-g)), but the procedure applies to any $\eta\geq 1/3$. For $\eta = 1$, the Dirac point is at $k = \pm \pi/2a$. Away from the inversion symmetric point, when $c_1 = c+m/2$ and $c_2 = c-m/2$, expanding the matrix $\mathbf{C}'(k=\pm\pi/2a+\delta k)$ for small $\delta k$ and $m$:
\begin{equation}
	\begin{split}
		\mathbf{C}'(\pm\pi/2a+\delta k) &\approx \begin{bmatrix}
			0 & 2(\pm 1 - i) ca \delta k - m (1 \pm i)/2\\
			2(\pm 1 + i) ca \delta k - m (1 \mp i)/2 & 0
		\end{bmatrix}\\
		&= -m(\boldsymbol{\upsigma}_1\mp\boldsymbol{\upsigma}_2)/2 + 2ca \delta k (\pm \boldsymbol{\upsigma}_1+\boldsymbol{\upsigma}_2).
	\end{split}
\end{equation}
As in the case of $k=\pi/a$, we create a domain wall at $x = 0$ with phase $c_1> c_2$ in the region $x<0$ and the phase $c_1<c_2$ in the region $x>0$, implying the position-dependence of mass $m(x)$, $i.e.$, $m(x)>0$ when $x<0$ and $m(x)<0$ when $x>0$. Since the translation symmetry is broken due to the presence of the domain wall, $\delta k \rightarrow -i \partial_x$. We seek a zero frequency domain wall eigenmode $\psi(x)$:
\begin{equation}
	\begin{split}
		& \mathbf{C}' \psi(x) = 0\\
		\Rightarrow & [-m(x)(\boldsymbol{\upsigma}_1\mp\boldsymbol{\upsigma}_2)/2 - 2ca i  (\pm \boldsymbol{\upsigma}_1+\boldsymbol{\upsigma}_2)\partial_x] \psi(x) = 0\\
		\Rightarrow & (\boldsymbol{\upsigma}_1\mp\boldsymbol{\upsigma}_2)[-m(x)(\boldsymbol{\upsigma}_1\mp\boldsymbol{\upsigma}_2)/2 - 2ca i  (\pm \boldsymbol{\upsigma}_1+\boldsymbol{\upsigma}_2)\partial_x] \psi(x) = 0\\ 
		\Rightarrow & [-m(x) \mathbf{1} + 4 c a \boldsymbol{\upsigma}_3 \partial_x] \psi(x) = 0.
	\end{split}
\end{equation}
Plugging the ansatz $\psi(x) = f(x)\begin{pmatrix} 1 \\0 \end{pmatrix}$, where $f(x)$ is a scalar function, we obtain the following differential equation for $f(x)$:
\begin{equation}
	\partial_x f(x) = \frac{m(x)}{4ca} \Rightarrow f(x) = c_0 e^{\frac{1}{4ca}\int_0^x dx' m(x')},
\end{equation}
where $c_0$ is a constant. Notice that $f(x)$ decays exponentially away from $x = 0$ since $\frac{m(x>0)}{4ca}<0$ and $\frac{m(x<0)}{4ca}>0$. Recalling that the zero frequency domain wall mode is at $k=\pm \pi/2a$, the full expression of the mode is:
\begin{equation}
	\psi_{\pm \pi/2a}(x) = \psi(x)e^{\pm i \pi x/2a} = c_0 e^{\frac{1}{4ca}\int_0^x dx' m(x')} e^{\pm i \pi x/2a} \begin{pmatrix} 1 \\0 \end{pmatrix}.
	\label{JR_other_sol}
\end{equation}
The SFT plots of the JR zero modes perfectly match the ones obtained from our supercell toy models presented in Fig.~\ref{JRmodes}, with Eqn.~\ref{JR_pi_smalleta} plotted in Figs.~\ref{JRmodes}a, and Fig. 
S4u in Supplementary Note 2, and Eqns.~\ref{JR_pi_largeeta} and ~\ref{JR_other_sol} in Figs.~\ref{JRmodes}b, S3, and S4v in Supplementary Note 2.
\subsubsection{Jackiw Rebbi Mode Parity and Hybridization}
In principle, the analytical solutions of JR modes in the SSH model all stay at $\omega/\omega_0=\sqrt{(c_1+c_2+2c')/m}$ [and $\omega^2=0$ if we plot the eigenvalues of $\mathbf{C}'(k)$]. This is because, to get the JR modes, an effective long-wavelength (around the Dirac point) approximation is used, and a domain wall is created by simply flipping the sign of $m$ from one side of the domain wall to the other. However, such a treatment overlooks the microscopic details at the domain wall of the actual system, $i.e.$, the way the two domains are connected. In the spring-mass system as shown in Fig.~\ref{supercell_toy}a, b, the TPDWSs are not at the mid-gap frequency ($i.e.$, $\omega/\omega_0=\sqrt{(c_1+c_2+2c')/m}$). There are two mechanisms behind the shift of their energies. First, to obtain the mid-gap TPDWS frequency, the dynamical matrix must present chiral symmetry, requiring the conservation of total spring constants for each mass, including the domain-wall mass. However, this is unachievable if one simply mirrors one side of the mass-spring chain about the domain wall since the spring constants about the domain-wall mass never equal ones about other masses, as shown in Fig.~\ref{supercell_toy}a, b, as well as in many other studies including only NNs, such as Fig. 3 in Chaunsali $et$  $al$.~\cite{chaunsali2017demonstrating}. Hence, the energy of the domain wall mode can shift up (down) from the mid-gap frequency if the domain wall mass is connected by stiff (soft) springs on both sides. Nonetheless, such a frequency shift can be avoided by adding an additional spring to the domain wall mass to achieve chiral symmetry, as presented in Fig. S4a, b in Supplementary Note 2. Their eigenvalues, $\omega^2/\omega_0^2$, after removing the diagonal elements using Eqn.~\ref{chiral_TNN}, are strictly symmetric about $\omega^2/\omega_0^2=0$, similar to the $zero-energy$ in electronic systems, as shown in Fig. S4c, d.

Second, even if we force the dynamical matrix to be chiral with an additional spring at the domain wall as in Fig. S4a, b, three JR modes in the case of strong TNNs can still hybridize with one another, shifting the TPDWSs from the mid-gap due to band hybridization dictated by parities of the JR modes. For example, in the case of SC1 with $c'=1$ shown in Fig. S4a, the JR mode due to the Dirac cone $k =\pi$ is parity odd under inversion. The other two JR modes are not eigenstates of the inversion operator. However, the hybridization of these modes creates a cosine and a sine function, the former of which is parity even and the latter parity odd. When combined with the one at $k=\pi$, we get two antisymmetric eigenmodes with a displacement amplitude of zero at the domain-wall mass ($i.e.$, two parity-odd modes) and one symmetric with a non-zero amplitude domain-wall mass displacement ($i.e.$, one parity-even mode). This is evident from the SC1 TPDWSs shown in blue in Fig. S4d, where two have zero amplitudes at the domain wall mass (because, in total, there are two parity-odd TPDWSs, $i.e.$, SC1-E1/3) shifted symmetrically up/down from $\omega^2/\omega_0^2=0$, while the other one, SC1-E2, located exactly at $\omega^2/\omega_0^2=0$ presents a nonzero domain-wall-mass amplitude (the parity-even mode). The scenario is the opposite for SC2, in which the JR mode at $k =\pi$ is even under inversion. Thus, there is only one asymmetric TPDW with a zero displacement amplitude at the domain wall mass ($i.e.$, one parity-odd mode, SC2-E2) located at $\omega^2/\omega_0^2=0$ and two symmetric ones with a non-zero-amplitude domain wall mass displacement ($i.e.$, two parity-even modes, SC2-E1/3) with frequencies shifted up/down symmetrically about $\omega^2/\omega_0^2=0$.

An important point to note here is that only JR modes of the $same$ parity can hybridize due to symmetry constraints. For example, in SC1 of Fig. S4a, the two odd modes hybridize, and one of the resulting hybridized modes shifts up in frequency, and the other one shifts down. Since the even mode cannot hybridize with the other two, it remains in the middle. Similar results also hold (albeit with parities flipped) for SC2. In either case, the JR mode at $k=\pi$ always ends up mixing with one of the other two modes with the same parity with frequency shifts. Thus, these two shifted TPDWSs always present a peak at $k=\pi$ in their SFT plots, as shown in Fig. S4v. On the other hand, the other mode at $\omega^2/\omega_0^2=0$ not mixed with the one at $k=\pi$ has SFT peaks only located at the other two Dirac points in the IBZ.

The shifting of the zero-frequency(energy) boundary/domain-wall modes (after removing the diagonal elements) to finite frequencies is not unique to the SSH model, instead, it is a generic feature that arises in most known 0D topological modes in various types of topological insulators (such as corner modes in 2D higher-order topological insulators), whose energies are also sensitive to local perturbations near the localized modes\cite{van2020topological,proctor2020robustness}.

\subsection{Topologically Protected Domain-Wall States beyond Equal Third-Nearest Neighbors}
The Berry connection proposed in this work provides a generalized paradigm to predict the number of TPDWSs and their wave properties. The equal TNN scenario discussed in Results and Discussion is an example when the winding number fails, while the Berry connection succeeds in predicting the number of TPDWSs. In cases when the effect of TNN difference dominates, as in Fig. S5a in Supplementary Note 3, the Berry connection makes the same prediction of the TPDWSs as the winding number does, however with additional wave information. For example, with $c_1'=3c-\Delta c'$ and $c_2'=3c+\Delta c'$, where $\Delta c'=0.1c$, the winding number difference between the two gauges presented in Fig. S5d is one, yet $\Delta B(k)$ shows two peaks and one valley from $k=0$ to $2\pi$, corresponding to three distinct edge modes similar to the scenario in Results and Discussion. With an enhanced $c'$ difference, such as when $\Delta c'=0.3c$, we obtain three distinct peaks in $\Delta B(k)$ in IBZ at $k=1.23$, $\pi$, and $5.05$, as presented in Fig. S5b, c, corresponding to the three Dirac points predicted with Eqn.~\ref{Dirac_points} with $\eta=3$. Integration of these three local peaks all yields one, suggesting three TPDWSs existing in the bulk bandgap. These three TPDWSs also coincide with the winding number of three due to the all positive signs of the local peaks of $\Delta B(k)$. The contour plot of the off-diagonal element of $\mathbf{C}'(k)$ expressed in Eqn.~\ref{chiral_TNN} also shows winding numbers of -1 and 2 for the two gauges presented in Fig. S5e, yielding a difference of 3 between the two phases. Thus, these three TPDWSs are expected using either the Berry connection or the winding number calculation. Their existence can be confirmed by conducting a supercell analysis in the same manner described in Supercell Analysis of the Su-Schrieffer-Heeger Model. The band diagram, mode shapes, and the SFT of the TPDWS and bulk modes are presented in Fig. S5f-l. As can be seen from the SFT plots in Fig. S5l, peak locations of the TPDWSs all match those in the $\Delta B(k)$ plot presented in Fig. S5b, indicating consistent TPDWS wavelengths predicted with Berry connections. The evolution of $\Delta B(k)$ with $\Delta c'$ also suggests the breakdown of the winding number prediction fails when $\Delta c'<0.2c$ due to the flip of the peak and valley in $\Delta B(k)$ at $k=\pi$ due to a winding number difference of one (shown in Fig. S5d) as opposed to three from both the $\Delta B(k)$ calculation and the supercell analysis.

Moreover, the Berry connection calculation is also applicable to lattices beyond TNN interactions. For example, with identical TNNs ($c'$) and fifth nearest neighbors (FNNs), $c''$, and nonidentical NNs ($i.e.$, $c_1 \neq c_2$), the winding number difference is still one, as shown in Fig. S6a in Supplementary Note 3. However, the Berry connection $\Delta B(k)$ reveals three peaks and two valleys with local integrals $\pm 1$, Fig. S6b, suggesting five TPDWSs existing in the bulk bandgap. Indeed, from the supercell analysis with a similar setup as shown in Figs.~\ref{supercell_toy}a and S5a, five edge modes emerge within the bulk bandgap, Fig. S6c, with their mode shapes presented in Fig. S6d-j. SFT of these five TPDWSs in Fig. S6k reveals that all of them are a hybridization of five wavelengths with wave numbers corresponding to the locations of peaks and valleys shown in Fig. S6b. One can prove in a similar fashion that as long as the differences in BNNs are sufficiently small, the winding numbers of two gauges will always yield zero and one, inconsistent with the actual number of TPDWSs, which can, nonetheless, be conveniently captured by the $\Delta B(k)$ calculation.

\subsection{Experimental Fabrication and Characterization}
The specimens are 3D-printed (Stratasys F170 FDM 3D Printer) using acrylonitrile butadiene styrene (ABS) with the following parameters: Young's modulus $E$=1.5 GPa, Poisson's ratio $\mu$=0.35, and density $\rho$=1250 kg $\mathrm{m}^{-3}$. As presented in Fig.~\ref{experiment}d, each unit cell contains a pair of masses (green cubes) with side length $W_m$=6 mm, connected by 5 mm-nearest-neighboring (NN) struts with alternating radii, $r_1$=3.52 mm (blue) and $r_2$=1.47 mm (yellow), to enable stiffer and softer NN interactions, respectively. Strong (weak) TNNs are established by a combination of red squared frames with side length $W$=16 mm, height $H$=4 mm (3.2 mm), and thickness $t$=1.33 mm (1.07 mm), and bars with radius $r_3$=2.43 mm (1.28 mm) connecting the masses and frames. Mode shapes of the three (one) edge modes and two bulk modes with strong (weak) TNNs modeled by COMSOL Multiphysics with these material properties and structural dimensions are presented in Fig. S7 in Supplementary Note 4. As we can see from Fig. S7a, the three TPDWSs in the lattice with strong TNNs, from high to low frequencies, are asymmetric, symmetric, and symmetric manners about the domain wall, while the one with weak TNNs, as shown in Fig. S7b is asymmetric about the domain wall. The Symmetries and locations of the deformed frames all match well with the ones obtained from experiments presented in Fig.~\ref{experiment}i, j.
\section{Data Availability}
The datasets generated during and/or analyzed during the current study are available from the corresponding author on reasonable request.
\vspace{0.2cm}
\section{Code Availability}
All MATLAB codes generated for the current study are available from the corresponding author on reasonable request and report of its use should cite this paper.
\vspace{0.2cm}


\begin{thebibliography}{59}
\ifx \bisbn   \undefined \def \bisbn  #1{ISBN #1}\fi
\ifx \binits  \undefined \def \binits#1{#1}\fi
\ifx \bauthor  \undefined \def \bauthor#1{#1}\fi
\ifx \batitle  \undefined \def \batitle#1{#1}\fi
\ifx \bjtitle  \undefined \def \bjtitle#1{#1}\fi
\ifx \bvolume  \undefined \def \bvolume#1{\textbf{#1}}\fi
\ifx \byear  \undefined \def \byear#1{#1}\fi
\ifx \bissue  \undefined \def \bissue#1{#1}\fi
\ifx \bfpage  \undefined \def \bfpage#1{#1}\fi
\ifx \blpage  \undefined \def \blpage #1{#1}\fi
\ifx \burl  \undefined \def \burl#1{\textsf{#1}}\fi
\ifx \doiurl  \undefined \def \doiurl#1{\url{https://doi.org/#1}}\fi
\ifx \betal  \undefined \def \betal{\textit{et al.}}\fi
\ifx \binstitute  \undefined \def \binstitute#1{#1}\fi
\ifx \binstitutionaled  \undefined \def \binstitutionaled#1{#1}\fi
\ifx \bctitle  \undefined \def \bctitle#1{#1}\fi
\ifx \beditor  \undefined \def \beditor#1{#1}\fi
\ifx \bpublisher  \undefined \def \bpublisher#1{#1}\fi
\ifx \bbtitle  \undefined \def \bbtitle#1{#1}\fi
\ifx \bedition  \undefined \def \bedition#1{#1}\fi
\ifx \bseriesno  \undefined \def \bseriesno#1{#1}\fi
\ifx \blocation  \undefined \def \blocation#1{#1}\fi
\ifx \bsertitle  \undefined \def \bsertitle#1{#1}\fi
\ifx \bsnm \undefined \def \bsnm#1{#1}\fi
\ifx \bsuffix \undefined \def \bsuffix#1{#1}\fi
\ifx \bparticle \undefined \def \bparticle#1{#1}\fi
\ifx \barticle \undefined \def \barticle#1{#1}\fi
\bibcommenthead
\ifx \bconfdate \undefined \def \bconfdate #1{#1}\fi
\ifx \botherref \undefined \def \botherref #1{#1}\fi
\ifx \url \undefined \def \url#1{\textsf{#1}}\fi
\ifx \bchapter \undefined \def \bchapter#1{#1}\fi
\ifx \bbook \undefined \def \bbook#1{#1}\fi
\ifx \bcomment \undefined \def \bcomment#1{#1}\fi
\ifx \oauthor \undefined \def \oauthor#1{#1}\fi
\ifx \citeauthoryear \undefined \def \citeauthoryear#1{#1}\fi
\ifx \endbibitem  \undefined \def \endbibitem {}\fi
\ifx \bconflocation  \undefined \def \bconflocation#1{#1}\fi
\ifx \arxivurl  \undefined \def \arxivurl#1{\textsf{#1}}\fi
\csname PreBibitemsHook\endcsname

\bibitem[\protect\citeauthoryear{Haldane}{1988}]{haldane1988model}
\begin{barticle}
\bauthor{\bsnm{Haldane}, \binits{F.D.M.}}:
\batitle{Model for a quantum hall effect without landau levels:
  Condensed-matter realization of the" parity anomaly"}.
\bjtitle{Physical Review Letters}
\bvolume{61}(\bissue{18}),
\bfpage{2015}
(\byear{1988})
\end{barticle}
\endbibitem

\bibitem[\protect\citeauthoryear{Kane and Mele}{2005}]{kane2005quantum}
\begin{barticle}
\bauthor{\bsnm{Kane}, \binits{C.L.}},
\bauthor{\bsnm{Mele}, \binits{E.J.}}:
\batitle{Quantum spin hall effect in graphene}.
\bjtitle{Physical review letters}
\bvolume{95}(\bissue{22}),
\bfpage{226801}
(\byear{2005})
\end{barticle}
\endbibitem

\bibitem[\protect\citeauthoryear{Hasan and Kane}{2010}]{hasan2010colloquium}
\begin{barticle}
\bauthor{\bsnm{Hasan}, \binits{M.Z.}},
\bauthor{\bsnm{Kane}, \binits{C.L.}}:
\batitle{Colloquium: topological insulators}.
\bjtitle{Reviews of Modern Physics}
\bvolume{82}(\bissue{4}),
\bfpage{3045}
(\byear{2010})
\end{barticle}
\endbibitem

\bibitem[\protect\citeauthoryear{Qi and Zhang}{2011}]{qi2011topological}
\begin{barticle}
\bauthor{\bsnm{Qi}, \binits{X.-L.}},
\bauthor{\bsnm{Zhang}, \binits{S.-C.}}:
\batitle{Topological insulators and superconductors}.
\bjtitle{Reviews of Modern Physics}
\bvolume{83}(\bissue{4}),
\bfpage{1057}
(\byear{2011})
\end{barticle}
\endbibitem

\bibitem[\protect\citeauthoryear{Kane and Lubensky}{2014}]{kane2014topological}
\begin{barticle}
\bauthor{\bsnm{Kane}, \binits{C.}},
\bauthor{\bsnm{Lubensky}, \binits{T.}}:
\batitle{Topological boundary modes in isostatic lattices}.
\bjtitle{Nature Physics}
\bvolume{10}(\bissue{1}),
\bfpage{39}
(\byear{2014})
\end{barticle}
\endbibitem

\bibitem[\protect\citeauthoryear{Paulose et~al.}{2015}]{paulose2015topological}
\begin{barticle}
\bauthor{\bsnm{Paulose}, \binits{J.}},
\bauthor{\bsnm{Chen}, \binits{B.G.-g.}},
\bauthor{\bsnm{Vitelli}, \binits{V.}}:
\batitle{Topological modes bound to dislocations in mechanical metamaterials}.
\bjtitle{Nature Physics}
\bvolume{11}(\bissue{2}),
\bfpage{153}
(\byear{2015})
\end{barticle}
\endbibitem

\bibitem[\protect\citeauthoryear{Rocklin
  et~al.}{2017}]{rocklin2017transformable}
\begin{barticle}
\bauthor{\bsnm{Rocklin}, \binits{D.Z.}},
\bauthor{\bsnm{Zhou}, \binits{S.}},
\bauthor{\bsnm{Sun}, \binits{K.}},
\bauthor{\bsnm{Mao}, \binits{X.}}:
\batitle{Transformable topological mechanical metamaterials}.
\bjtitle{Nature communications}
\bvolume{8},
\bfpage{14201}
(\byear{2017})
\end{barticle}
\endbibitem

\bibitem[\protect\citeauthoryear{Rocklin et~al.}{2016}]{rocklin2016mechanical}
\begin{barticle}
\bauthor{\bsnm{Rocklin}, \binits{D.Z.}},
\bauthor{\bsnm{Chen}, \binits{B.G.-g.}},
\bauthor{\bsnm{Falk}, \binits{M.}},
\bauthor{\bsnm{Vitelli}, \binits{V.}},
\bauthor{\bsnm{Lubensky}, \binits{T.}}:
\batitle{Mechanical weyl modes in topological maxwell lattices}.
\bjtitle{Physical review letters}
\bvolume{116}(\bissue{13}),
\bfpage{135503}
(\byear{2016})
\end{barticle}
\endbibitem

\bibitem[\protect\citeauthoryear{Stenull et~al.}{2016}]{stenull2016topological}
\begin{barticle}
\bauthor{\bsnm{Stenull}, \binits{O.}},
\bauthor{\bsnm{Kane}, \binits{C.}},
\bauthor{\bsnm{Lubensky}, \binits{T.}}:
\batitle{Topological phonons and weyl lines in three dimensions}.
\bjtitle{Physical review letters}
\bvolume{117}(\bissue{6}),
\bfpage{068001}
(\byear{2016})
\end{barticle}
\endbibitem

\bibitem[\protect\citeauthoryear{Bilal et~al.}{2017}]{bilal2017intrinsically}
\begin{botherref}
\oauthor{\bsnm{Bilal}, \binits{O.R.}},
\oauthor{\bsnm{S{\"u}sstrunk}, \binits{R.}},
\oauthor{\bsnm{Daraio}, \binits{C.}},
\oauthor{\bsnm{Huber}, \binits{S.D.}}:
Intrinsically polar elastic metamaterials.
Advanced Materials
\textbf{29}(26)
(2017)
\end{botherref}
\endbibitem

\bibitem[\protect\citeauthoryear{Ma et~al.}{2018}]{ma2018edge}
\begin{barticle}
\bauthor{\bsnm{Ma}, \binits{J.}},
\bauthor{\bsnm{Zhou}, \binits{D.}},
\bauthor{\bsnm{Sun}, \binits{K.}},
\bauthor{\bsnm{Mao}, \binits{X.}},
\bauthor{\bsnm{Gonella}, \binits{S.}}:
\batitle{Edge modes and asymmetric wave transport in topological lattices:
  Experimental characterization at finite frequencies}.
\bjtitle{Physical review letters}
\bvolume{121}(\bissue{9}),
\bfpage{094301}
(\byear{2018})
\end{barticle}
\endbibitem

\bibitem[\protect\citeauthoryear{Ma et~al.}{2019}]{ma2019valley}
\begin{barticle}
\bauthor{\bsnm{Ma}, \binits{J.}},
\bauthor{\bsnm{Sun}, \binits{K.}},
\bauthor{\bsnm{Gonella}, \binits{S.}}:
\batitle{Valley hall in-plane edge states as building blocks for elastodynamic
  logic circuits}.
\bjtitle{Physical Review Applied}
\bvolume{12}(\bissue{4}),
\bfpage{044015}
(\byear{2019})
\end{barticle}
\endbibitem

\bibitem[\protect\citeauthoryear{S{\"u}sstrunk and
  Huber}{2015}]{susstrunk2015observation}
\begin{barticle}
\bauthor{\bsnm{S{\"u}sstrunk}, \binits{R.}},
\bauthor{\bsnm{Huber}, \binits{S.D.}}:
\batitle{Observation of phononic helical edge states in a mechanical
  topological insulator}.
\bjtitle{Science}
\bvolume{349}(\bissue{6243}),
\bfpage{47}--\blpage{50}
(\byear{2015})
\end{barticle}
\endbibitem

\bibitem[\protect\citeauthoryear{Nash et~al.}{2015}]{nash2015topological}
\begin{barticle}
\bauthor{\bsnm{Nash}, \binits{L.M.}},
\bauthor{\bsnm{Kleckner}, \binits{D.}},
\bauthor{\bsnm{Read}, \binits{A.}},
\bauthor{\bsnm{Vitelli}, \binits{V.}},
\bauthor{\bsnm{Turner}, \binits{A.M.}},
\bauthor{\bsnm{Irvine}, \binits{W.T.}}:
\batitle{Topological mechanics of gyroscopic metamaterials}.
\bjtitle{Proceedings of the National Academy of Sciences}
\bvolume{112}(\bissue{47}),
\bfpage{14495}--\blpage{14500}
(\byear{2015})
\end{barticle}
\endbibitem

\bibitem[\protect\citeauthoryear{Wang et~al.}{2015}]{wang2015topological}
\begin{barticle}
\bauthor{\bsnm{Wang}, \binits{P.}},
\bauthor{\bsnm{Lu}, \binits{L.}},
\bauthor{\bsnm{Bertoldi}, \binits{K.}}:
\batitle{Topological phononic crystals with one-way elastic edge waves}.
\bjtitle{Physical review letters}
\bvolume{115}(\bissue{10}),
\bfpage{104302}
(\byear{2015})
\end{barticle}
\endbibitem

\bibitem[\protect\citeauthoryear{Mousavi
  et~al.}{2015}]{mousavi2015topologically}
\begin{barticle}
\bauthor{\bsnm{Mousavi}, \binits{S.H.}},
\bauthor{\bsnm{Khanikaev}, \binits{A.B.}},
\bauthor{\bsnm{Wang}, \binits{Z.}}:
\batitle{Topologically protected elastic waves in phononic metamaterials}.
\bjtitle{Nature communications}
\bvolume{6},
\bfpage{8682}
(\byear{2015})
\end{barticle}
\endbibitem

\bibitem[\protect\citeauthoryear{Kariyado and
  Hatsugai}{2015}]{kariyado2015manipulation}
\begin{barticle}
\bauthor{\bsnm{Kariyado}, \binits{T.}},
\bauthor{\bsnm{Hatsugai}, \binits{Y.}}:
\batitle{Manipulation of dirac cones in mechanical graphene}.
\bjtitle{Scientific reports}
\bvolume{5},
\bfpage{18107}
(\byear{2015})
\end{barticle}
\endbibitem

\bibitem[\protect\citeauthoryear{Pal et~al.}{2016}]{pal2016helical}
\begin{barticle}
\bauthor{\bsnm{Pal}, \binits{R.K.}},
\bauthor{\bsnm{Schaeffer}, \binits{M.}},
\bauthor{\bsnm{Ruzzene}, \binits{M.}}:
\batitle{Helical edge states and topological phase transitions in phononic
  systems using bi-layered lattices}.
\bjtitle{Journal of Applied Physics}
\bvolume{119}(\bissue{8}),
\bfpage{084305}
(\byear{2016})
\end{barticle}
\endbibitem

\bibitem[\protect\citeauthoryear{Brendel
  et~al.}{2017}]{brendel2017pseudomagnetic}
\begin{barticle}
\bauthor{\bsnm{Brendel}, \binits{C.}},
\bauthor{\bsnm{Peano}, \binits{V.}},
\bauthor{\bsnm{Painter}, \binits{O.J.}},
\bauthor{\bsnm{Marquardt}, \binits{F.}}:
\batitle{Pseudomagnetic fields for sound at the nanoscale}.
\bjtitle{Proceedings of the National Academy of Sciences}
\bvolume{114}(\bissue{17}),
\bfpage{3390}--\blpage{3395}
(\byear{2017})
\end{barticle}
\endbibitem

\bibitem[\protect\citeauthoryear{Chaunsali
  et~al.}{2018}]{chaunsali2018subwavelength}
\begin{barticle}
\bauthor{\bsnm{Chaunsali}, \binits{R.}},
\bauthor{\bsnm{Chen}, \binits{C.-W.}},
\bauthor{\bsnm{Yang}, \binits{J.}}:
\batitle{Subwavelength and directional control of flexural waves in
  zone-folding induced topological plates}.
\bjtitle{Physical Review B}
\bvolume{97}(\bissue{5}),
\bfpage{054307}
(\byear{2018})
\end{barticle}
\endbibitem

\bibitem[\protect\citeauthoryear{Prodan et~al.}{2017}]{prodan2017dynamical}
\begin{barticle}
\bauthor{\bsnm{Prodan}, \binits{E.}},
\bauthor{\bsnm{Dobiszewski}, \binits{K.}},
\bauthor{\bsnm{Kanwal}, \binits{A.}},
\bauthor{\bsnm{Palmieri}, \binits{J.}},
\bauthor{\bsnm{Prodan}, \binits{C.}}:
\batitle{Dynamical majorana edge modes in a broad class of topological
  mechanical systems}.
\bjtitle{Nature communications}
\bvolume{8},
\bfpage{14587}
(\byear{2017})
\end{barticle}
\endbibitem

\bibitem[\protect\citeauthoryear{Luo et~al.}{2021}]{luo2021observation}
\begin{barticle}
\bauthor{\bsnm{Luo}, \binits{L.}},
\bauthor{\bsnm{Wang}, \binits{H.-X.}},
\bauthor{\bsnm{Lin}, \binits{Z.-K.}},
\bauthor{\bsnm{Jiang}, \binits{B.}},
\bauthor{\bsnm{Wu}, \binits{Y.}},
\bauthor{\bsnm{Li}, \binits{F.}},
\bauthor{\bsnm{Jiang}, \binits{J.-H.}}:
\batitle{Observation of a phononic higher-order weyl semimetal}.
\bjtitle{Nature Materials}
\bvolume{20}(\bissue{6}),
\bfpage{794}--\blpage{799}
(\byear{2021})
\end{barticle}
\endbibitem

\bibitem[\protect\citeauthoryear{Wang and Wei}{2021}]{wang2021elastic}
\begin{barticle}
\bauthor{\bsnm{Wang}, \binits{Z.}},
\bauthor{\bsnm{Wei}, \binits{Q.}}:
\batitle{An elastic higher-order topological insulator based on kagome phononic
  crystals}.
\bjtitle{Journal of Applied Physics}
\bvolume{129}(\bissue{3}),
\bfpage{035102}
(\byear{2021})
\end{barticle}
\endbibitem

\bibitem[\protect\citeauthoryear{Ni et~al.}{2019}]{ni2019observation}
\begin{barticle}
\bauthor{\bsnm{Ni}, \binits{X.}},
\bauthor{\bsnm{Weiner}, \binits{M.}},
\bauthor{\bsnm{Alu}, \binits{A.}},
\bauthor{\bsnm{Khanikaev}, \binits{A.B.}}:
\batitle{Observation of higher-order topological acoustic states protected by
  generalized chiral symmetry}.
\bjtitle{Nature materials}
\bvolume{18}(\bissue{2}),
\bfpage{113}--\blpage{120}
(\byear{2019})
\end{barticle}
\endbibitem

\bibitem[\protect\citeauthoryear{Qi et~al.}{2020}]{qi2020acoustic}
\begin{barticle}
\bauthor{\bsnm{Qi}, \binits{Y.}},
\bauthor{\bsnm{Qiu}, \binits{C.}},
\bauthor{\bsnm{Xiao}, \binits{M.}},
\bauthor{\bsnm{He}, \binits{H.}},
\bauthor{\bsnm{Ke}, \binits{M.}},
\bauthor{\bsnm{Liu}, \binits{Z.}}:
\batitle{Acoustic realization of quadrupole topological insulators}.
\bjtitle{Physical Review Letters}
\bvolume{124}(\bissue{20}),
\bfpage{206601}
(\byear{2020})
\end{barticle}
\endbibitem

\bibitem[\protect\citeauthoryear{Zhang et~al.}{2019}]{zhang2019second}
\begin{barticle}
\bauthor{\bsnm{Zhang}, \binits{X.}},
\bauthor{\bsnm{Wang}, \binits{H.-X.}},
\bauthor{\bsnm{Lin}, \binits{Z.-K.}},
\bauthor{\bsnm{Tian}, \binits{Y.}},
\bauthor{\bsnm{Xie}, \binits{B.}},
\bauthor{\bsnm{Lu}, \binits{M.-H.}},
\bauthor{\bsnm{Chen}, \binits{Y.-F.}},
\bauthor{\bsnm{Jiang}, \binits{J.-H.}}:
\batitle{Second-order topology and multidimensional topological transitions in
  sonic crystals}.
\bjtitle{Nature Physics}
\bvolume{15}(\bissue{6}),
\bfpage{582}--\blpage{588}
(\byear{2019})
\end{barticle}
\endbibitem

\bibitem[\protect\citeauthoryear{Chen et~al.}{2021}]{chen2021corner}
\begin{barticle}
\bauthor{\bsnm{Chen}, \binits{C.-W.}},
\bauthor{\bsnm{Chaunsali}, \binits{R.}},
\bauthor{\bsnm{Christensen}, \binits{J.}},
\bauthor{\bsnm{Theocharis}, \binits{G.}},
\bauthor{\bsnm{Yang}, \binits{J.}}:
\batitle{Corner states in a second-order mechanical topological insulator}.
\bjtitle{Communications Materials}
\bvolume{2}(\bissue{1}),
\bfpage{1}--\blpage{6}
(\byear{2021})
\end{barticle}
\endbibitem

\bibitem[\protect\citeauthoryear{Su et~al.}{1979}]{su1979solitons}
\begin{barticle}
\bauthor{\bsnm{Su}, \binits{W.}},
\bauthor{\bsnm{Schrieffer}, \binits{J.}},
\bauthor{\bsnm{Heeger}, \binits{A.J.}}:
\batitle{Solitons in polyacetylene}.
\bjtitle{Physical review letters}
\bvolume{42}(\bissue{25}),
\bfpage{1698}
(\byear{1979})
\end{barticle}
\endbibitem

\bibitem[\protect\citeauthoryear{Su et~al.}{1980}]{su1980soliton}
\begin{barticle}
\bauthor{\bsnm{Su}, \binits{W.-P.}},
\bauthor{\bsnm{Schrieffer}, \binits{J.}},
\bauthor{\bsnm{Heeger}, \binits{A.}}:
\batitle{Soliton excitations in polyacetylene}.
\bjtitle{Physical Review B}
\bvolume{22}(\bissue{4}),
\bfpage{2099}
(\byear{1980})
\end{barticle}
\endbibitem

\bibitem[\protect\citeauthoryear{Lubensky et~al.}{2015}]{lubensky2015phonons}
\begin{barticle}
\bauthor{\bsnm{Lubensky}, \binits{T.}},
\bauthor{\bsnm{Kane}, \binits{C.}},
\bauthor{\bsnm{Mao}, \binits{X.}},
\bauthor{\bsnm{Souslov}, \binits{A.}},
\bauthor{\bsnm{Sun}, \binits{K.}}:
\batitle{Phonons and elasticity in critically coordinated lattices}.
\bjtitle{Reports on Progress in Physics}
\bvolume{78}(\bissue{7}),
\bfpage{073901}
(\byear{2015})
\end{barticle}
\endbibitem

\bibitem[\protect\citeauthoryear{Esmann et~al.}{2018}]{esmann2018topological}
\begin{barticle}
\bauthor{\bsnm{Esmann}, \binits{M.}},
\bauthor{\bsnm{Lamberti}, \binits{F.}},
\bauthor{\bsnm{Lema{\^\i}tre}, \binits{A.}},
\bauthor{\bsnm{Lanzillotti-Kimura}, \binits{N.}}:
\batitle{Topological acoustics in coupled nanocavity arrays}.
\bjtitle{Physical Review B}
\bvolume{98}(\bissue{16}),
\bfpage{161109}
(\byear{2018})
\end{barticle}
\endbibitem

\bibitem[\protect\citeauthoryear{Pal and Ruzzene}{2017}]{pal2017edge}
\begin{barticle}
\bauthor{\bsnm{Pal}, \binits{R.K.}},
\bauthor{\bsnm{Ruzzene}, \binits{M.}}:
\batitle{Edge waves in plates with resonators: an elastic analogue of the
  quantum valley hall effect}.
\bjtitle{New Journal of Physics}
\bvolume{19}(\bissue{2}),
\bfpage{025001}
(\byear{2017})
\end{barticle}
\endbibitem

\bibitem[\protect\citeauthoryear{Zak}{1989}]{zak1989berry}
\begin{barticle}
\bauthor{\bsnm{Zak}, \binits{J.}}:
\batitle{Berry’s phase for energy bands in solids}.
\bjtitle{Physical review letters}
\bvolume{62}(\bissue{23}),
\bfpage{2747}
(\byear{1989})
\end{barticle}
\endbibitem

\bibitem[\protect\citeauthoryear{Lu et~al.}{2016}]{lu2016valley}
\begin{barticle}
\bauthor{\bsnm{Lu}, \binits{J.}},
\bauthor{\bsnm{Qiu}, \binits{C.}},
\bauthor{\bsnm{Ke}, \binits{M.}},
\bauthor{\bsnm{Liu}, \binits{Z.}}:
\batitle{Valley vortex states in sonic crystals}.
\bjtitle{Physical review letters}
\bvolume{116}(\bissue{9}),
\bfpage{093901}
(\byear{2016})
\end{barticle}
\endbibitem

\bibitem[\protect\citeauthoryear{Lu et~al.}{2017}]{lu2017observation}
\begin{barticle}
\bauthor{\bsnm{Lu}, \binits{J.}},
\bauthor{\bsnm{Qiu}, \binits{C.}},
\bauthor{\bsnm{Ye}, \binits{L.}},
\bauthor{\bsnm{Fan}, \binits{X.}},
\bauthor{\bsnm{Ke}, \binits{M.}},
\bauthor{\bsnm{Zhang}, \binits{F.}},
\bauthor{\bsnm{Liu}, \binits{Z.}}:
\batitle{Observation of topological valley transport of sound in sonic
  crystals}.
\bjtitle{Nature Physics}
\bvolume{13}(\bissue{4}),
\bfpage{369}--\blpage{374}
(\byear{2017})
\end{barticle}
\endbibitem

\bibitem[\protect\citeauthoryear{Liu and Semperlotti}{2018}]{liu2018tunable}
\begin{barticle}
\bauthor{\bsnm{Liu}, \binits{T.-W.}},
\bauthor{\bsnm{Semperlotti}, \binits{F.}}:
\batitle{Tunable acoustic valley--hall edge states in reconfigurable phononic
  elastic waveguides}.
\bjtitle{Physical Review Applied}
\bvolume{9}(\bissue{1}),
\bfpage{014001}
(\byear{2018})
\end{barticle}
\endbibitem

\bibitem[\protect\citeauthoryear{Liu and
  Semperlotti}{2019}]{liu2019experimental}
\begin{barticle}
\bauthor{\bsnm{Liu}, \binits{T.-W.}},
\bauthor{\bsnm{Semperlotti}, \binits{F.}}:
\batitle{Experimental evidence of robust acoustic valley hall edge states in a
  nonresonant topological elastic waveguide}.
\bjtitle{Physical Review Applied}
\bvolume{11}(\bissue{1}),
\bfpage{014040}
(\byear{2019})
\end{barticle}
\endbibitem

\bibitem[\protect\citeauthoryear{Chen et~al.}{2021}]{chen2021roton}
\begin{barticle}
\bauthor{\bsnm{Chen}, \binits{Y.}},
\bauthor{\bsnm{Kadic}, \binits{M.}},
\bauthor{\bsnm{Wegener}, \binits{M.}}:
\batitle{Roton-like acoustical dispersion relations in 3d metamaterials}.
\bjtitle{Nature communications}
\bvolume{12}(\bissue{1}),
\bfpage{1}--\blpage{8}
(\byear{2021})
\end{barticle}
\endbibitem

\bibitem[\protect\citeauthoryear{Iglesias~Mart{\'\i}nez
  et~al.}{2021}]{iglesias2021experimental}
\begin{barticle}
\bauthor{\bsnm{Iglesias~Mart{\'\i}nez}, \binits{J.A.}},
\bauthor{\bsnm{Gro{\ss}}, \binits{M.F.}},
\bauthor{\bsnm{Chen}, \binits{Y.}},
\bauthor{\bsnm{Frenzel}, \binits{T.}},
\bauthor{\bsnm{Laude}, \binits{V.}},
\bauthor{\bsnm{Kadic}, \binits{M.}},
\bauthor{\bsnm{Wegener}, \binits{M.}}:
\batitle{Experimental observation of roton-like dispersion relations in
  metamaterials}.
\bjtitle{Science Advances}
\bvolume{7}(\bissue{49}),
\bfpage{2189}
(\byear{2021})
\end{barticle}
\endbibitem

\bibitem[\protect\citeauthoryear{Iorio et~al.}{2022}]{iorio2022roton}
\begin{botherref}
\oauthor{\bsnm{Iorio}, \binits{L.}},
\oauthor{\bsnm{De~Ponti}, \binits{J.M.}},
\oauthor{\bsnm{Maspero}, \binits{F.}},
\oauthor{\bsnm{Ardito}, \binits{R.}}:
Roton-like dispersion via polarisation change for elastic wave energy control
  in graded delay-lines.
arXiv preprint arXiv:2211.09431
(2022)
\end{botherref}
\endbibitem

\bibitem[\protect\citeauthoryear{Cui et~al.}{2022}]{cui2022tunable}
\begin{barticle}
\bauthor{\bsnm{Cui}, \binits{J.-G.}},
\bauthor{\bsnm{Yang}, \binits{T.}},
\bauthor{\bsnm{Niu}, \binits{M.-Q.}},
\bauthor{\bsnm{Chen}, \binits{L.-Q.}}:
\batitle{Tunable roton-like dispersion relation with parametric excitations}.
\bjtitle{Journal of Applied Mechanics}
\bvolume{89}(\bissue{11}),
\bfpage{111005}
(\byear{2022})
\end{barticle}
\endbibitem

\bibitem[\protect\citeauthoryear{Wang et~al.}{2022}]{wang2022nonlocal}
\begin{barticle}
\bauthor{\bsnm{Wang}, \binits{K.}},
\bauthor{\bsnm{Chen}, \binits{Y.}},
\bauthor{\bsnm{Kadic}, \binits{M.}},
\bauthor{\bsnm{Wang}, \binits{C.}},
\bauthor{\bsnm{Wegener}, \binits{M.}}:
\batitle{Nonlocal interaction engineering of 2d roton-like dispersion relations
  in acoustic and mechanical metamaterials}.
\bjtitle{Communications Materials}
\bvolume{3}(\bissue{1}),
\bfpage{1}--\blpage{11}
(\byear{2022})
\end{barticle}
\endbibitem

\bibitem[\protect\citeauthoryear{Zhu et~al.}{2022}]{zhu2022observation}
\begin{barticle}
\bauthor{\bsnm{Zhu}, \binits{Z.}},
\bauthor{\bsnm{Gao}, \binits{Z.}},
\bauthor{\bsnm{Liu}, \binits{G.-G.}},
\bauthor{\bsnm{Ge}, \binits{Y.}},
\bauthor{\bsnm{Wang}, \binits{Y.}},
\bauthor{\bsnm{Xi}, \binits{X.}},
\bauthor{\bsnm{Yan}, \binits{B.}},
\bauthor{\bsnm{Chen}, \binits{F.}},
\bauthor{\bsnm{Shum}, \binits{P.P.}},
\bauthor{\bsnm{Sun}, \binits{H.-x.}}, \betal:
\batitle{Observation of multiple rotons and multidirectional roton-like
  dispersion relations in acoustic metamaterials}.
\bjtitle{New Journal of Physics}
\bvolume{24}(\bissue{12}),
\bfpage{123019}
(\byear{2022})
\end{barticle}
\endbibitem

\bibitem[\protect\citeauthoryear{Grundmann}{2020}]{grundmann2020topological}
\begin{barticle}
\bauthor{\bsnm{Grundmann}, \binits{M.}}:
\batitle{Topological states due to third-neighbor coupling in diatomic linear
  elastic chains}.
\bjtitle{physica status solidi (b)}
\bvolume{257}(\bissue{9}),
\bfpage{2000176}
(\byear{2020})
\end{barticle}
\endbibitem

\bibitem[\protect\citeauthoryear{Chen et~al.}{2018}]{chen2018study}
\begin{barticle}
\bauthor{\bsnm{Chen}, \binits{H.}},
\bauthor{\bsnm{Nassar}, \binits{H.}},
\bauthor{\bsnm{Huang}, \binits{G.}}:
\batitle{A study of topological effects in 1d and 2d mechanical lattices}.
\bjtitle{Journal of the Mechanics and Physics of Solids}
\bvolume{117},
\bfpage{22}--\blpage{36}
(\byear{2018})
\end{barticle}
\endbibitem

\bibitem[\protect\citeauthoryear{Liu et~al.}{2023}]{liu2023acoustic}
\begin{barticle}
\bauthor{\bsnm{Liu}, \binits{H.}},
\bauthor{\bsnm{Huang}, \binits{X.}},
\bauthor{\bsnm{Yan}, \binits{M.}},
\bauthor{\bsnm{Lu}, \binits{J.}},
\bauthor{\bsnm{Deng}, \binits{W.}},
\bauthor{\bsnm{Liu}, \binits{Z.}}:
\batitle{Acoustic topological metamaterials of large winding number}.
\bjtitle{Physical Review Applied}
\bvolume{19}(\bissue{5}),
\bfpage{054028}
(\byear{2023})
\end{barticle}
\endbibitem

\bibitem[\protect\citeauthoryear{Jackiw and Rebbi}{1976}]{jackiw1976solitons}
\begin{barticle}
\bauthor{\bsnm{Jackiw}, \binits{R.}},
\bauthor{\bsnm{Rebbi}, \binits{C.}}:
\batitle{Solitons with fermion number $1/2$}.
\bjtitle{Physical Review D}
\bvolume{13}(\bissue{12}),
\bfpage{3398}
(\byear{1976})
\end{barticle}
\endbibitem

\bibitem[\protect\citeauthoryear{Guzm{\'a}n et~al.}{2022}]{guzman2022geometry}
\begin{barticle}
\bauthor{\bsnm{Guzm{\'a}n}, \binits{M.}},
\bauthor{\bsnm{Bartolo}, \binits{D.}},
\bauthor{\bsnm{Carpentier}, \binits{D.}}:
\batitle{Geometry and topology tango in ordered and amorphous chiral matter}.
\bjtitle{SciPost Physics}
\bvolume{12}(\bissue{1}),
\bfpage{038}
(\byear{2022})
\end{barticle}
\endbibitem

\bibitem[\protect\citeauthoryear{van Miert and
  Ortix}{2020}]{van2020topological}
\begin{barticle}
\bauthor{\bsnm{Miert}, \binits{G.}},
\bauthor{\bsnm{Ortix}, \binits{C.}}:
\batitle{On the topological immunity of corner states in two-dimensional
  crystalline insulators}.
\bjtitle{npj Quantum Materials}
\bvolume{5}(\bissue{1}),
\bfpage{63}
(\byear{2020})
\end{barticle}
\endbibitem

\bibitem[\protect\citeauthoryear{Proctor et~al.}{2020}]{proctor2020robustness}
\begin{barticle}
\bauthor{\bsnm{Proctor}, \binits{M.}},
\bauthor{\bsnm{Huidobro}, \binits{P.A.}},
\bauthor{\bsnm{Bradlyn}, \binits{B.}},
\bauthor{\bsnm{De~Paz}, \binits{M.B.}},
\bauthor{\bsnm{Vergniory}, \binits{M.G.}},
\bauthor{\bsnm{Bercioux}, \binits{D.}},
\bauthor{\bsnm{Garc{\'\i}a-Etxarri}, \binits{A.}}:
\batitle{Robustness of topological corner modes in photonic crystals}.
\bjtitle{Physical Review Research}
\bvolume{2}(\bissue{4}),
\bfpage{042038}
(\byear{2020})
\end{barticle}
\endbibitem

\bibitem[\protect\citeauthoryear{Sun et~al.}{2012}]{sun2012surface}
\begin{barticle}
\bauthor{\bsnm{Sun}, \binits{K.}},
\bauthor{\bsnm{Souslov}, \binits{A.}},
\bauthor{\bsnm{Mao}, \binits{X.}},
\bauthor{\bsnm{Lubensky}, \binits{T.}}:
\batitle{Surface phonons, elastic response, and conformal invariance in twisted
  kagome lattices}.
\bjtitle{Proceedings of the National Academy of Sciences}
\bvolume{109}(\bissue{31}),
\bfpage{12369}--\blpage{12374}
(\byear{2012})
\end{barticle}
\endbibitem

\bibitem[\protect\citeauthoryear{Prodan and
  Prodan}{2009}]{prodan2009topological}
\begin{barticle}
\bauthor{\bsnm{Prodan}, \binits{E.}},
\bauthor{\bsnm{Prodan}, \binits{C.}}:
\batitle{Topological phonon modes and their role in dynamic instability of
  microtubules}.
\bjtitle{Physical review letters}
\bvolume{103}(\bissue{24}),
\bfpage{248101}
(\byear{2009})
\end{barticle}
\endbibitem

\bibitem[\protect\citeauthoryear{Ma}{2023}]{ma2023phonon}
\begin{barticle}
\bauthor{\bsnm{Ma}, \binits{J.}}:
\batitle{Phonon engineering of micro-and nanophononic crystals and acoustic
  metamaterials: A review}.
\bjtitle{Small Science}
\bvolume{3}(\bissue{1}),
\bfpage{2200052}
(\byear{2023})
\end{barticle}
\endbibitem

\bibitem[\protect\citeauthoryear{Vanden-Hehir et~al.}{2019}]{vanden2019raman}
\begin{barticle}
\bauthor{\bsnm{Vanden-Hehir}, \binits{S.}},
\bauthor{\bsnm{Tipping}, \binits{W.J.}},
\bauthor{\bsnm{Lee}, \binits{M.}},
\bauthor{\bsnm{Brunton}, \binits{V.G.}},
\bauthor{\bsnm{Williams}, \binits{A.}},
\bauthor{\bsnm{Hulme}, \binits{A.N.}}:
\batitle{Raman imaging of nanocarriers for drug delivery}.
\bjtitle{Nanomaterials}
\bvolume{9}(\bissue{3}),
\bfpage{341}
(\byear{2019})
\end{barticle}
\endbibitem

\bibitem[\protect\citeauthoryear{Yeo et~al.}{2010}]{yeo2010ultrasonic}
\begin{barticle}
\bauthor{\bsnm{Yeo}, \binits{L.Y.}},
\bauthor{\bsnm{Friend}, \binits{J.R.}},
\bauthor{\bsnm{McIntosh}, \binits{M.P.}},
\bauthor{\bsnm{Meeusen}, \binits{E.N.}},
\bauthor{\bsnm{Morton}, \binits{D.A.}}:
\batitle{Ultrasonic nebulization platforms for pulmonary drug delivery}.
\bjtitle{Expert opinion on drug delivery}
\bvolume{7}(\bissue{6}),
\bfpage{663}--\blpage{679}
(\byear{2010})
\end{barticle}
\endbibitem

\bibitem[\protect\citeauthoryear{Bienfait et~al.}{2019}]{bienfait2019phonon}
\begin{barticle}
\bauthor{\bsnm{Bienfait}, \binits{A.}},
\bauthor{\bsnm{Satzinger}, \binits{K.J.}},
\bauthor{\bsnm{Zhong}, \binits{Y.}},
\bauthor{\bsnm{Chang}, \binits{H.-S.}},
\bauthor{\bsnm{Chou}, \binits{M.-H.}},
\bauthor{\bsnm{Conner}, \binits{C.R.}},
\bauthor{\bsnm{Dumur}, \binits{{\'E}.}},
\bauthor{\bsnm{Grebel}, \binits{J.}},
\bauthor{\bsnm{Peairs}, \binits{G.A.}},
\bauthor{\bsnm{Povey}, \binits{R.G.}}, \betal:
\batitle{Phonon-mediated quantum state transfer and remote qubit entanglement}.
\bjtitle{Science}
\bvolume{364}(\bissue{6438}),
\bfpage{368}--\blpage{371}
(\byear{2019})
\end{barticle}
\endbibitem

\bibitem[\protect\citeauthoryear{Chen et~al.}{2020}]{chen2020entanglement}
\begin{barticle}
\bauthor{\bsnm{Chen}, \binits{J.}},
\bauthor{\bsnm{Rossi}, \binits{M.}},
\bauthor{\bsnm{Mason}, \binits{D.}},
\bauthor{\bsnm{Schliesser}, \binits{A.}}:
\batitle{Entanglement of propagating optical modes via a mechanical interface}.
\bjtitle{Nature Communications}
\bvolume{11}(\bissue{1}),
\bfpage{943}
(\byear{2020})
\end{barticle}
\endbibitem

\bibitem[\protect\citeauthoryear{Wang et~al.}{2019}]{wang2019hexagonal}
\begin{barticle}
\bauthor{\bsnm{Wang}, \binits{Y.}},
\bauthor{\bsnm{Lee}, \binits{J.}},
\bauthor{\bsnm{Zheng}, \binits{X.-Q.}},
\bauthor{\bsnm{Xie}, \binits{Y.}},
\bauthor{\bsnm{Feng}, \binits{P.X.-L.}}:
\batitle{Hexagonal boron nitride phononic crystal waveguides}.
\bjtitle{ACS Photonics}
\bvolume{6}(\bissue{12}),
\bfpage{3225}--\blpage{3232}
(\byear{2019})
\end{barticle}
\endbibitem

\bibitem[\protect\citeauthoryear{Chaunsali
  et~al.}{2017}]{chaunsali2017demonstrating}
\begin{barticle}
\bauthor{\bsnm{Chaunsali}, \binits{R.}},
\bauthor{\bsnm{Kim}, \binits{E.}},
\bauthor{\bsnm{Thakkar}, \binits{A.}},
\bauthor{\bsnm{Kevrekidis}, \binits{P.G.}},
\bauthor{\bsnm{Yang}, \binits{J.}}:
\batitle{Demonstrating an in situ topological band transition in cylindrical
  granular chains}.
\bjtitle{Physical Review Letters}
\bvolume{119}(\bissue{2}),
\bfpage{024301}
(\byear{2017})
\end{barticle}
\endbibitem

\end{thebibliography}

\section{Acknowledgement}
A.R.A. and J.M. thank the University of Vermont for faculty start-up funds. S.S. and K.S. acknowledge the Office of Naval Research (MURI N00014-20-1-2479) and K.S. acknowledges the National Science Foundation (NSF Grant No. PHY-1748958) for the support of this research.

\vspace{0.2cm}
\section{Author Contributions}
A.R.A. conducted numerical simulation, sample fabrication, and experimental characterization. S.S. and K.S. contributed to the theoretical analysis. J.M. conceived and supervised the project. All authors contributed to discussions and writing the manuscript.
\vspace{0.2cm} 
\section{Competing Interests}
The authors declare no competing interests.

\begin{center}
	\textbf{\large Supplemental Material: Breakdown of Conventional Winding Number Calculation in Lattices with Interactions Beyond Nearest Neighbors}
\end{center}
\setcounter{equation}{0}
\setcounter{figure}{0}
\setcounter{table}{0}
\setcounter{page}{1}
\setcounter{section}{0}
\makeatletter
\renewcommand{\theequation}{S\arabic{equation}}
\renewcommand{\thefigure}{S\arabic{figure}}
\renewcommand{\citenumfont}[1]{S#1}
\renewcommand{\bibnumfmt}[1]{[S#1]}

	\section{Supplementary Note 1: One-Dimensional Su-Schrieffer-Heeger Model}\label{sec1}

\begin{figure}[h!]
	\centering
	\includegraphics[scale=0.45]{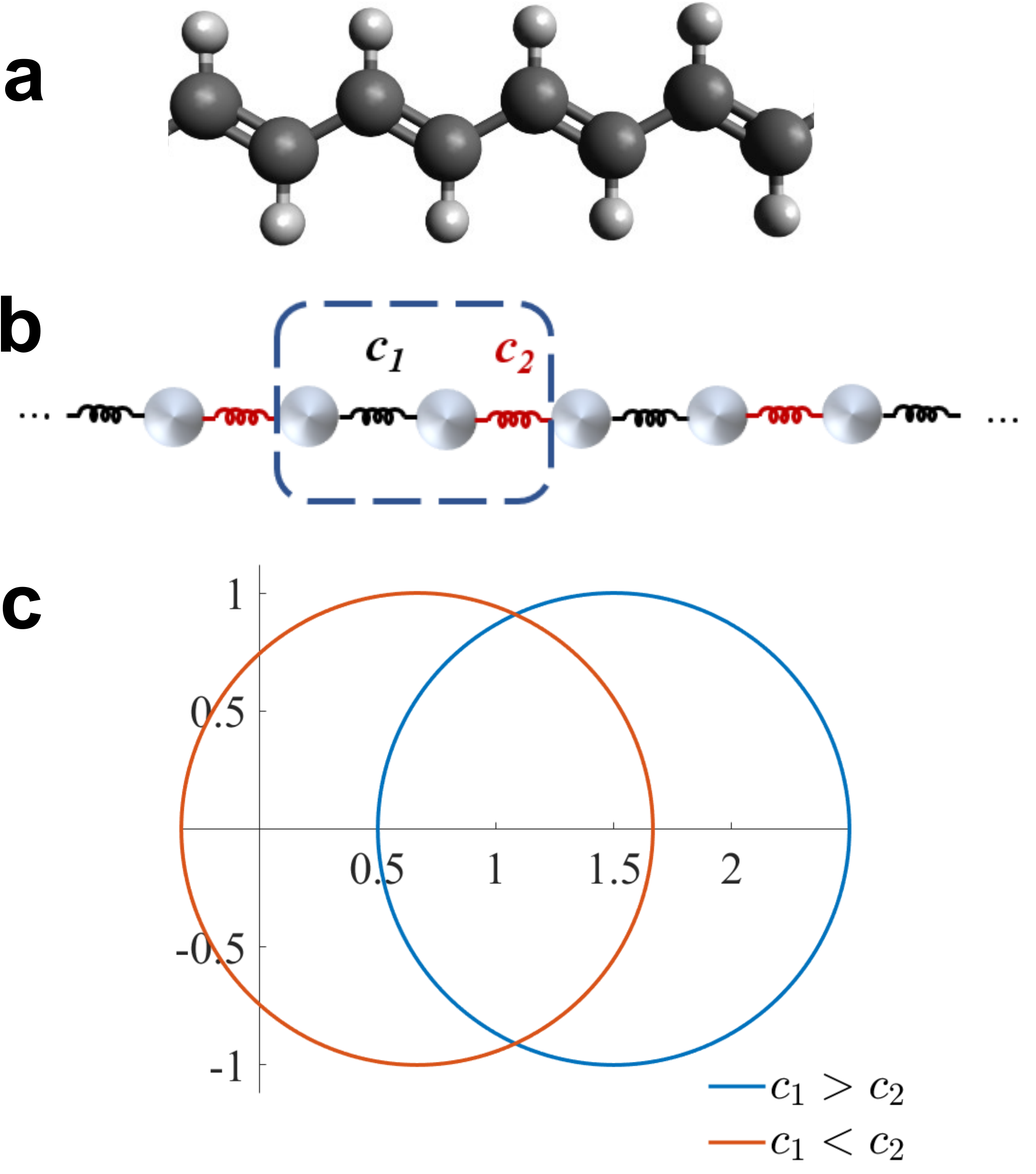} 
	\caption{\textbf{a} A polyacetylene chain created with Avogadro. \textbf{b} The simplified Su-Schrieffer-Heeger model consists of identical masses connected by alternating spring constants, $c_1$ and $c_2$, with the unit cell circled in a dashed line. \textbf{c} Contour plots in the complex plane of $\rho(k)$ for a complete circuit of $k$ from $k=0$ to $2\pi$. }
	\label{supp_1}
\end{figure}

\begin{figure}[h!]
	\centering
	\includegraphics[scale=0.45]{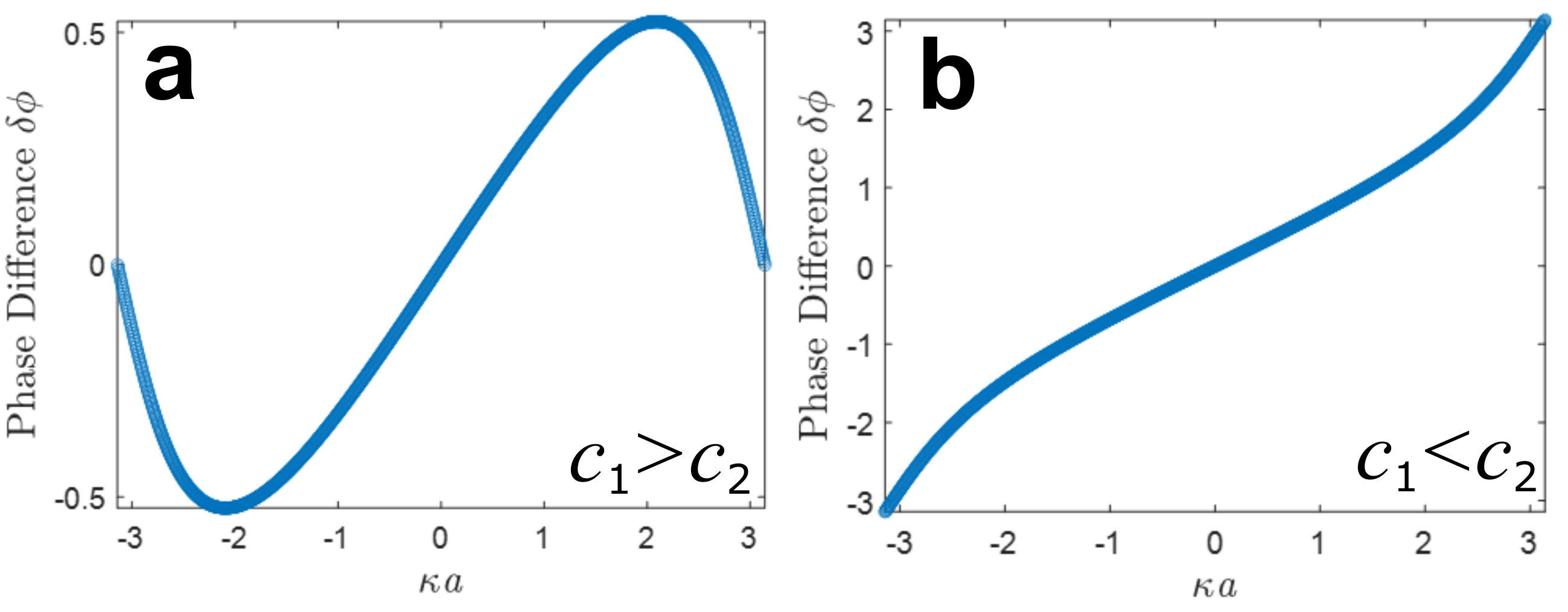} 
	\caption{Eigenvector phase difference across IBZ for cases of \textbf{a} $c_1>c_2$ and \textbf{b} $c_1<c_2$. }
	\label{supp_2}
\end{figure}

The one-dimensional (1D) Su-Schrieffer-Heeger (SSH) model was initially introduced to understand solitons in polyacetylene, shown in Fig. \ref{supp_1}a, which can be simplified as a 1D chain of identical masses, $m$, connected by alternating springs $c_1$ and $c_2$, Fig. \ref{supp_1}b. Gauge-dependent winding paths of the unit cell are shown in Fig. \ref{supp_1}c. Zak phases due to different gauge choice are presented in Fig. \ref{supp_2}.
\section{Supplementary Note 2: Supercell Analysis of the Su-Schrieffer-Heeger Model}
\begin{figure}[h!]
	\includegraphics[scale=0.35]{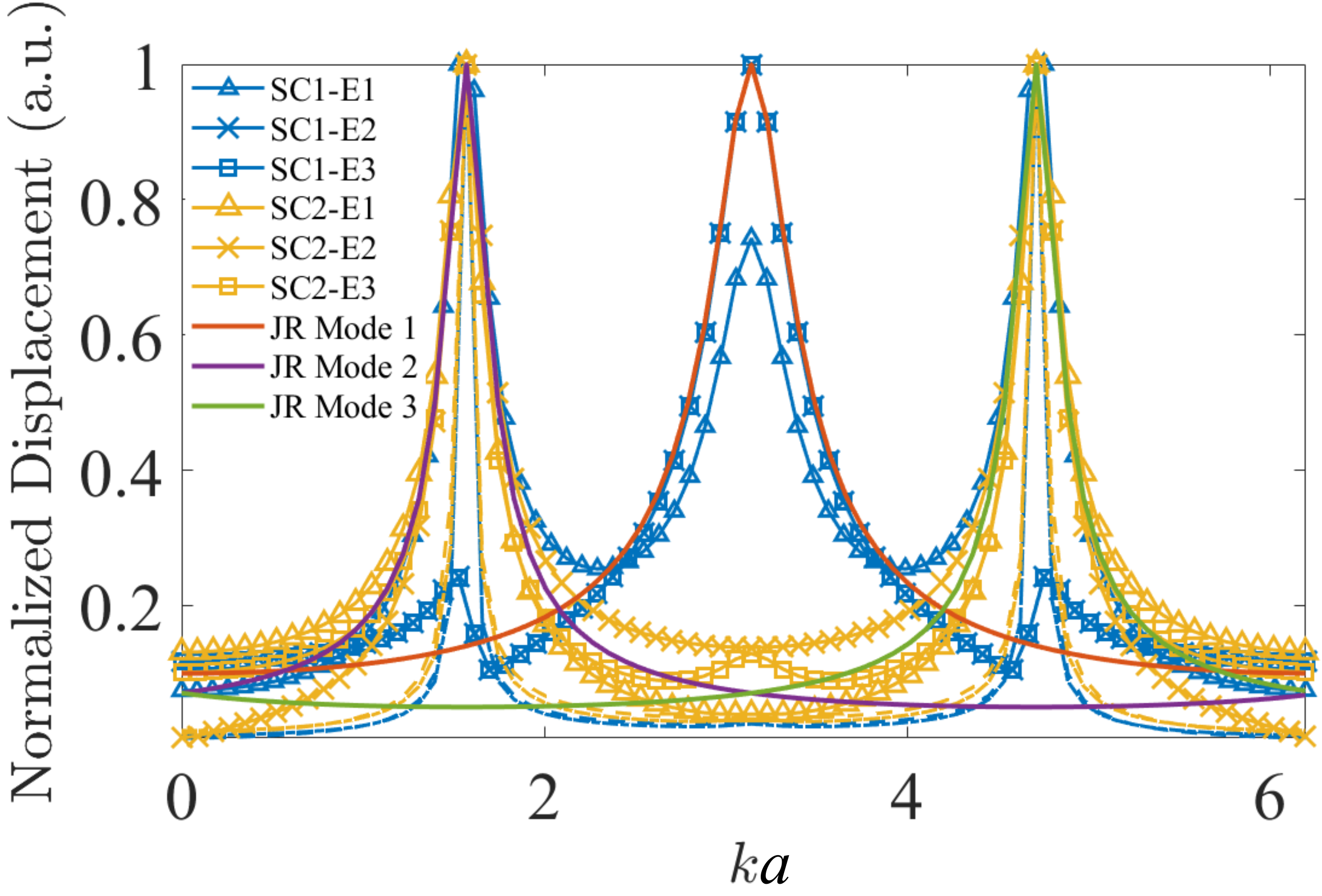}
	\centering
	\caption{Spatial Fourier transform of mode shapes in Fig. 2k-t in the main text starting from the mass right next to the domain wall: k (blue triangle), l (blue x), m (blue square), n (blue dot-dash), o (blue dash), p (yellow triangle), q (yellow x), r (yellow square), s (yellow dot-dash), and t (yellow dash)Solid curves are the Jackiw Rebbi zero modes.}
	\label{fourier_even}
\end{figure}

Spatial Fourier transform of mode shapes in Fig. 2k-t in the main text starting from the mass right next to the
domain wall are shown in Fig. \ref{fourier_even}.

\begin{figure}[h!]
	\centering
	\includegraphics[scale=0.1]{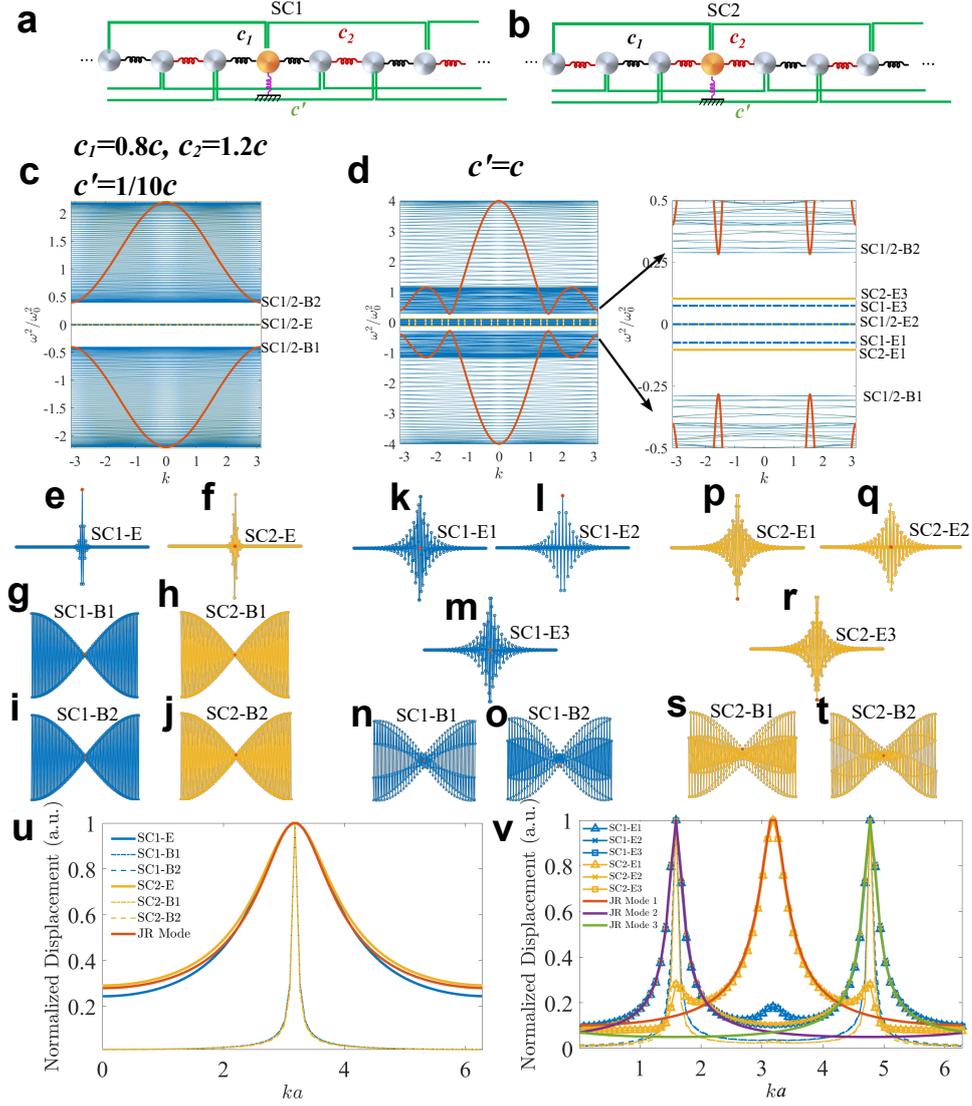} 
	\caption{\textbf{a} and \textbf{b} Supercells with the two arrangements of nearest neighbors with domain-wall mass connected to a fixed wall. \textbf{c} and \textbf{d}: $\omega^2/\omega_0^2$ of with \textbf{c} $c'=1/10c$ and \textbf{d} $c'=c$. Blue and yellow bands correspond to supercell \textbf{a} and \textbf{b}, respectively. Red curves are from the unit cell analysis. Dashed blue and bold solid yellow bands are edge modes, denoted as SC1-E(1-3) and SC2-E(1-3), respectively. Corresponding mode shapes are presented in \textbf{e}-\textbf{j} and \textbf{k}-\textbf{t}. \textbf{u} and \textbf{v}: Spatial Fourier transform of mode shapes. Solid curves are the Jackiw Rebbi (JR) zero modes.}
	\label{supp_supercell}
\end{figure}

To get the strict chiral stiffness matrix of the supercell, $\mathbf{C}(k)$, we add additional springs fixed to the ground to the domain-wall mass. Eigenvalues and eigen modes of the chiral matrix are presented in Fig. \ref{supp_supercell}.

\section{Supplementary Note 3: Topologically protected Domain-Wall States beyond Equal Third-Nearest Neighbors}
\begin{figure}[h!]
	\includegraphics[scale=0.16]{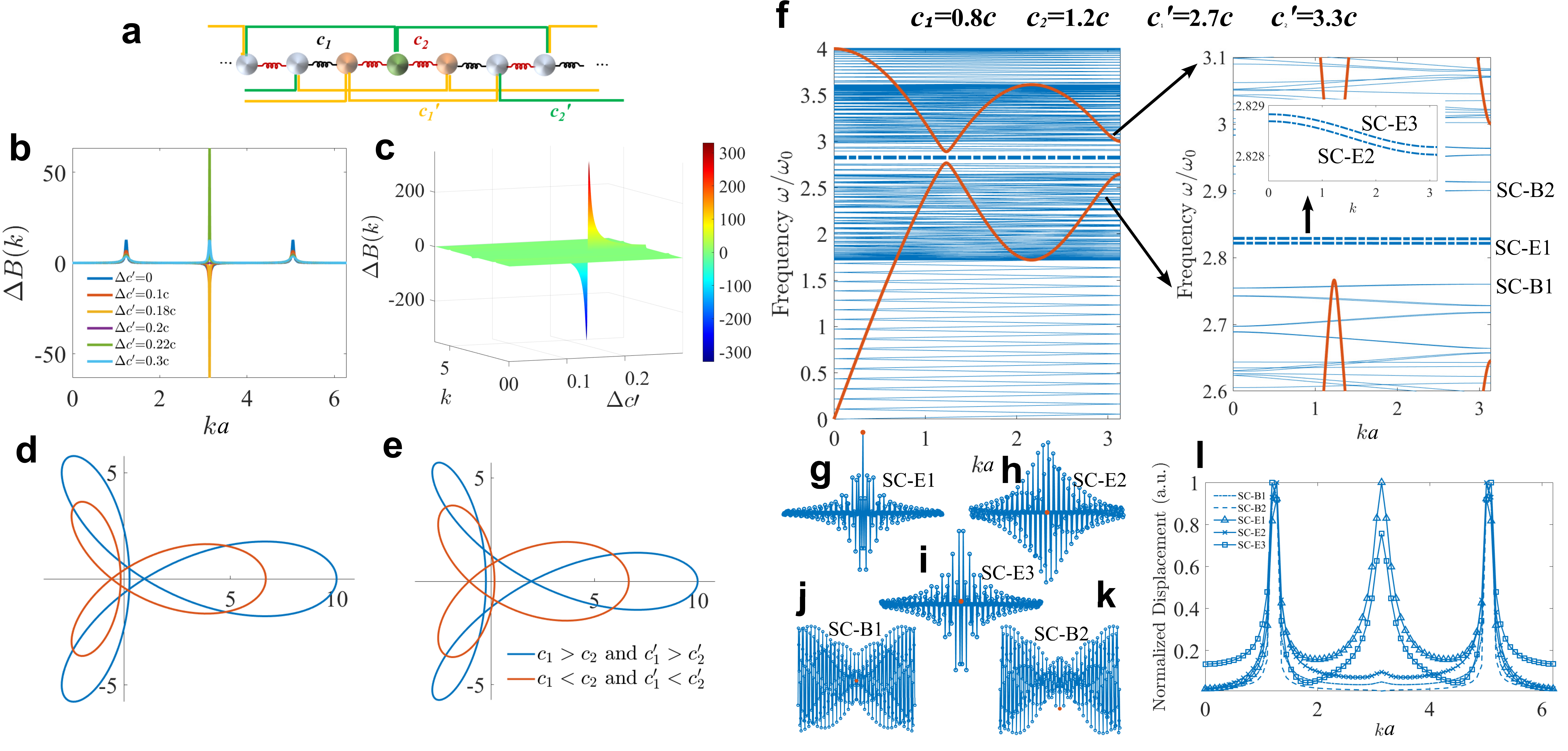} 
	\centering
	\caption{\textbf{a} A supercell featuring unequal nearest neighbors with $c_1=0.8c$ and $c_2=1.2c$ with stiff springs ($c_2$) connected to the green domain-wall mass, and nonidentical third nearest neighbors, $c_1'=3c-\Delta c'$ and $c_2'=3c+\Delta c'$, with strong neighbors ($c_2'$) connected to the green domain-wall mass and weak ($c_1'$) neighbors connected to the yellow domain-wall masses. \textbf{b} 2D and \textbf{c} 3D visualization of $\Delta B(k)$ from $k=0$ to $2\pi$ with different $\Delta c'$. \textbf{d} and \textbf{e} Contour plots of the off-diagonal element of the unit-cell stiffness matrix $\mathbf{C}'(k)$ in Eqn. 13 in the main text in the complex plane from $k=0$ to $2\pi$ for \textbf{d} $\Delta c'=0.1c$ and \textbf{e} $\Delta c'=0.3c$, respectively. \textbf{f} Band diagram (blue curves) of the supercell with spring constants listed above. Zoomed in are the three edge states (dashed blue curves) within the bulk bandgap (with bulk bands shown in red).  Mode shapes of these domain-wall modes are presented in \textbf{g}-\textbf{i} with labels of SC-E(1-3). To distinguish the edge modes from the bulk ones, bulk mode shapes for bands below and above the bandgap are also plotted in \textbf{j} and \textbf{k} with labels of SC-B1/2. Red solid circles in these mode-shape plots denote the displacements of the green domain-wall mass. Although visualized in the vertical directions, all mass displacements are $de$ $facto$ in the horizontal direction. Presented in \textbf{l} is the spatial Fourier transform of these mode shapes.}
	\label{unequal_cp}
\end{figure}

\begin{figure}[h!]
	\includegraphics[scale=0.22]{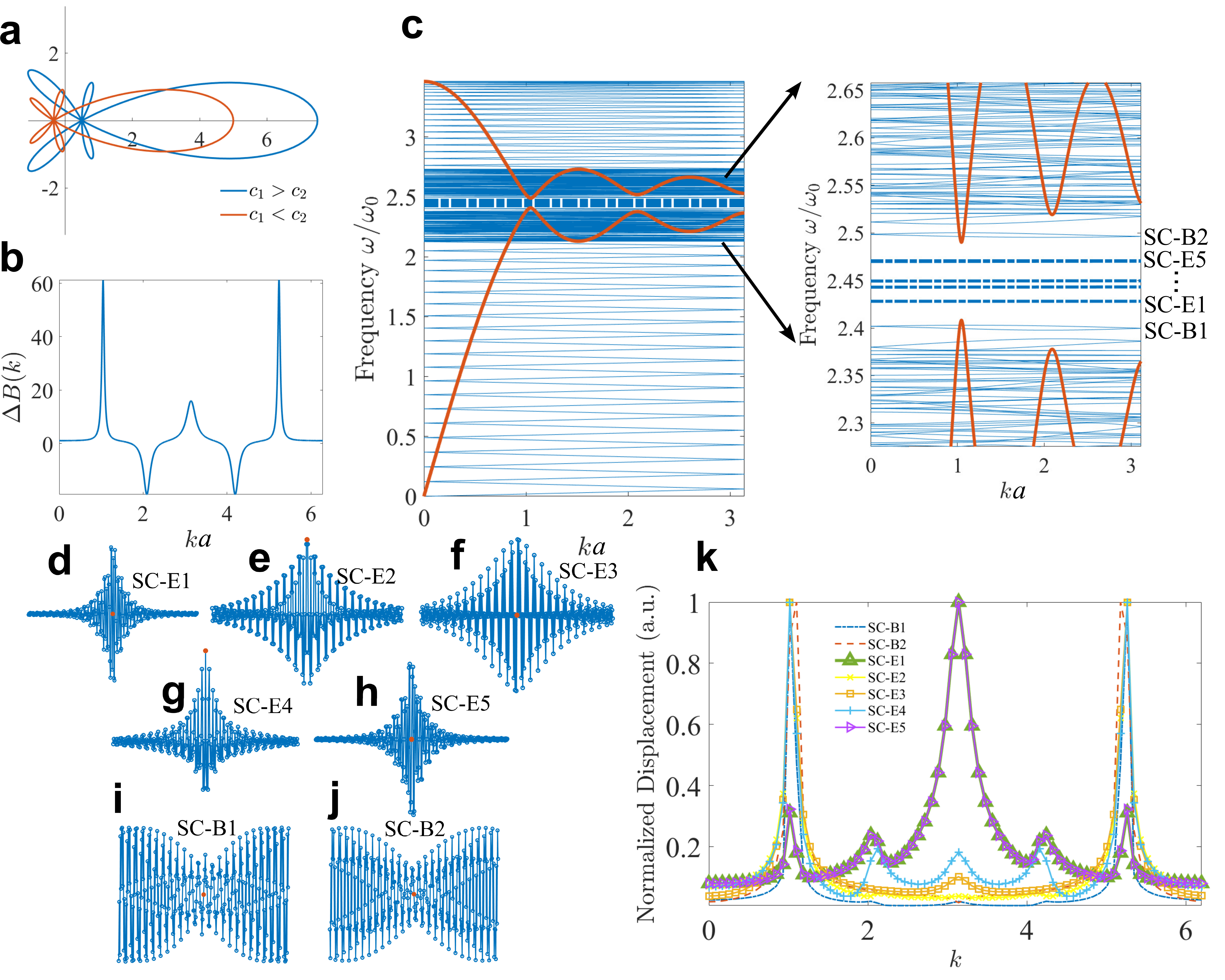} 
	\centering
	\caption{\textbf{a} Contour plots of the off-diagonal element of the stiffness matrix $\mathbf{C}'(k)$ in Eqn. 13 in the main text in the complex plane and \textbf{b} $\Delta B(k)$ from $k=0$ to $2\pi$ for a lattice with identical third ($c'=c$) and fifth ($c''=c$) nearest neighbors and nonidentical nearest neighbors $c_1=1.2c$ and $c_2=0.8c$. \textbf{c} Band diagram (blue curves) of the supercell with spring constants listed in the figure. Zoomed in are the three edge states (dashed blue curves) within the bulk bandgap (with bulk bands shown in red). Mode shapes of these edge and bulk modes are presented in \textbf{d}-\textbf{j} with labels of SC-E(1-3) and SC-B1/2, respectively.  Red solid circles in these mode-shape plots denote the displacements of the domain-wall mass. Although visualized in the vertical directions, all mass displacements are $de$ $facto$ in the horizontal direction. Presented in \textbf{k} is the spatial Fourier transform of these mode shapes. }
	\label{5NN}
\end{figure}

Lattices with unequal third-nearest neighbors and with fifth nearest neighbors are presented in Fig. \ref{unequal_cp} and Fig. \ref{5NN}, respectively.

\section{Supplementary Note 4: Finite Element Simulation of the Experimental Specimens}
\begin{figure}[h!]
	\includegraphics[scale=0.5]{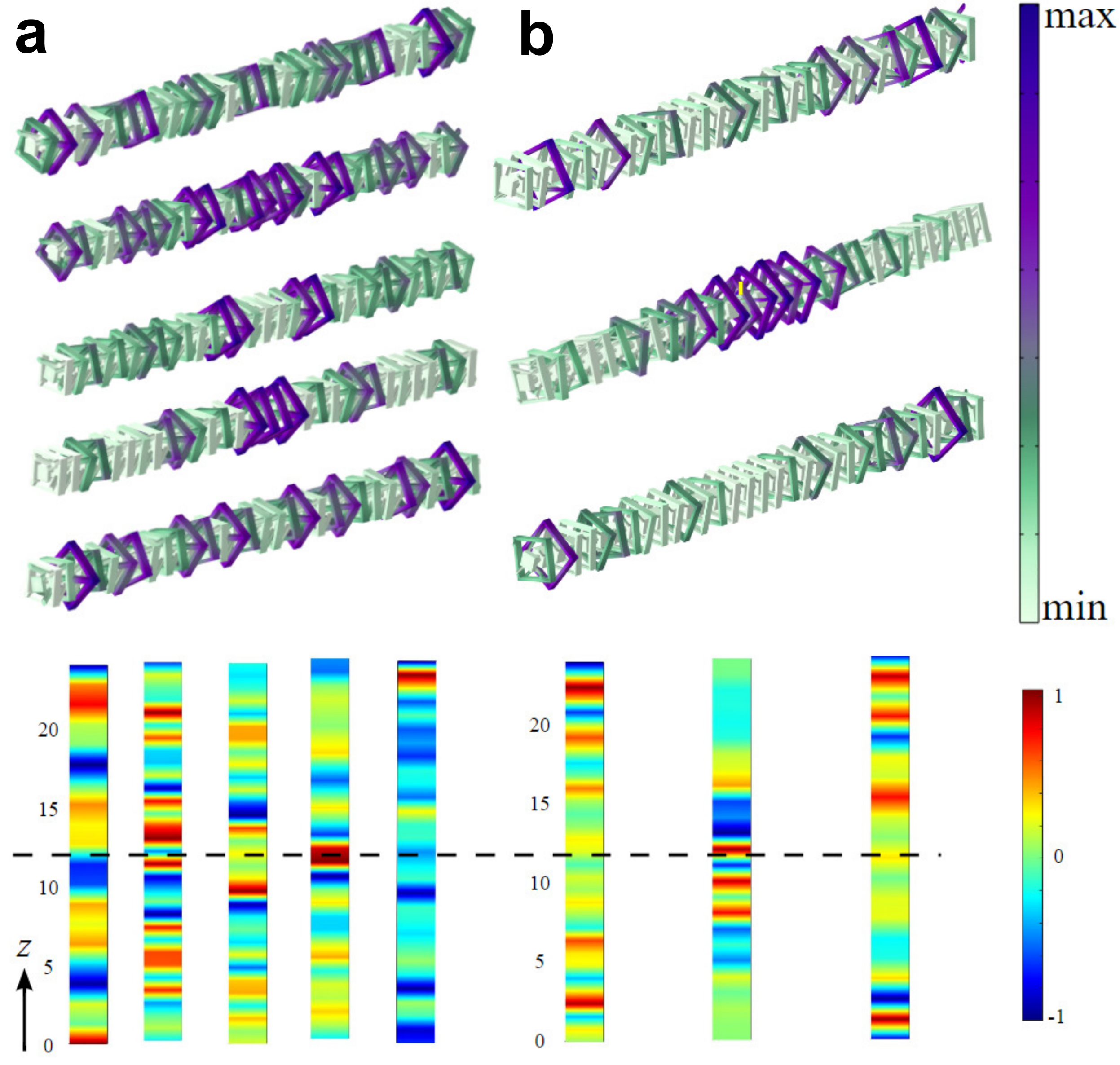}
	\centering
	\caption{3D visualization of the torsional displacements from the supercell analysis of the lattice with \textbf{a} strong and \textbf{b} weak third nearest neighbors.The green-purple color bar shows the magnitude of displacements. The top and bottom mode shapes in each column are the two bulk modes above and below the bandgap. The ones in between are the topologically protected domain wall states. The color bars below, from left to right, correspond to the torsional amplitudes of each frame of the lattices from top to bottom with decreasing frequencies, whose sequence also matches those in Fig. 5i and j in the main text, respectively. The dashed line labels the locations of the domain wall.} 
	\label{supp_3}
\end{figure}

We conduct a finite element analysis of the experimental specimens using COMSOL Multiphysics to compare with experimental measurement. The simulated mode shapes are presented in Fig. \ref{supp_3}. to compare with the ones obtained from experiments shown in Fig. 5i, j in the main text.
\end{document}